\documentclass[a4paper,11pt]{article}
\pdfoutput=1 
\usepackage{verbatim}

\usepackage{jheppub}
\usepackage{amsmath,amsfonts,amssymb,amsthm,graphicx,subfigure,color}
\usepackage{slashbox}
\usepackage[T1]{fontenc} 
\usepackage{enumerate}
\usepackage{tikz, tikz-3dplot, pgfplots}
\usepackage[utf8]{inputenc}
\usepackage{slashed}

\newcommand{\Vol}{{\rm vol}}

\title{\boldmath 	Holographic duals of five-dimensional SCFTs on a Riemann surface}

\author[a]{Ibrahima Bah,}
\author[b]{Achilleas Passias}
\author[a]{and Peter Weck} 
\affiliation[a]{Department of Physics and Astronomy, Johns Hopkins University, 3400 North Charles Street, Baltimore, MD 21218, USA}
\affiliation[b]{Department of Physics and Astronomy, Uppsala University,
Box 516, SE-75120 Uppsala, Sweden}

\emailAdd{iboubah@jhu.edu, achilleas.passias@physics.uu.se, pweck1@jhu.edu}

\abstract{We study the twisted compactifications of five-dimensional Seiberg SCFTs, with $SU_\mathcal{M}(2)\times E_{N_f+1}$ flavor symmetry, on a generic Riemann surface that preserves four supercharges.  The five-dimensional SCFTs are obtained from the decoupling limit of $N$ D4-branes probing a geometry of $N_f<8$ D8-branes and an O8-plane.  In addition to the R-symmetry, we can also twist the flavor symmetry by turning on background flux on the Riemann surface.  In particular, in the string theory construction of the five-dimensional SCFTs, the background flux for the $SU_\mathcal{M}(2)$ has a geometric origin, similar to the topological twist of the R-symmetry.  We argue that the resulting low-energy three-dimensional theories describe the dynamics on the world-volume of the $N$ D4-branes wrapped on the Riemann surface in the O8/D8 background.  The Riemann surface can be described as a curve in a Calabi-Yau three-fold that is a sum of two line bundles over it.  This allows for an explicit construction of $AdS_4$ solutions in massive IIA supergravity dual to the world-volume theories, thereby providing strong evidence that the three-dimensional SCFTs exist in the low-energy limit of the compactification of the five-dimensional SCFTs.  We compute observables such as the free energy and the scaling dimensions of operators dual to D2-brane probes; these have non-trivial dependence on the twist parameter for the $U(1)$ in $SU_\mathcal{M}(2)$.  The free energy exhibits the $N^{5/2}$ scaling that is emblematic of five-dimensional SCFTs.  
}
\preprint{UUITP-28/18}

\begin{document} 

\maketitle
\flushbottom

\newpage

%%%%%%%%%%%%%%%%%%%%%%%%%%%%%%%%%%%%%%%%%%%
\section{Introduction}

Indeed, one of the most useful ways to study Superconformal Field Theories (SCFTs) has been to consider twisted compactifications of the six-dimensional $(2,0)$ SCFT on various types of manifolds.  This approach has been championed by the class $\mathcal{S}$ program in the study of four-dimensional $\mathcal{N}=2$ SCFTs \cite{Gaiotto:2009we,Gaiotto:2009gz}.  The central idea has been to turn the identification of SCFTs in various dimensions into a problem of classifying geometries.  The various properties of SCFTs can be extracted from topology and geometry.  

The class $\mathcal{S}$ program and its generalizations have used six-dimensional SCFTs since they live in the highest dimension where superconformal algebras can exist \cite{Nahm:1977tg}.  In this sense the $(2,0)$ SCFTs behave as the origin for lower dimensional physics.  They are obtained by considering various decoupling limits of strings theories.

In five dimensions, there also exist a class of SCFTs with eight supercharges that are obtained by considering the decoupling limit of $N$ D4-branes in a background of $N_f<8$ D8-branes with an O8-plane \cite{Seiberg:1996bd} -- dubbed Seiberg SCFTs (reviewed in section \ref{fivescft}).\footnote{See also \cite{Intriligator:1997pq} for further studies and generalizations of the Seiberg SCFTs.}  These SCFTs carry an $SU_\mathcal{M}(2) \times E_{N_f+1}$ flavor symmetry.  It is natural to consider whether the Seiberg theories can be used to study, in a geometric way, lower dimensional SCFTs.  In this paper we initiate such a program by considering twisted compactifications of the Seiberg theories on a generic Riemann surface, possibly with punctures.  This is a problem that can be studied in holography, where we classify the possible $AdS_4$ solutions in massive IIA supergravity with suitable constraints imposed by the necessary conditions for supersymmetric compactifications of five-dimensional SCFTs on a Riemann surface.  

We restrict to cases where the resulting three-dimensional quantum field theory preserves at least four supercharges; these are given by topological twists of the five-dimensional SCFTs \cite{Witten:1988ze,Bershadsky:1995qy} on the Riemann surface.  This question is exactly analogous to the one studied for the holographic duals of $\mathcal{N}=1$ class $\mathcal{S}$ theories and their $(1,0)$ cousins in \cite{Benini:2009mz,Bah:2011vv,Bah:2012dg,Bah:2013qya,Bah:2015fwa,Apruzzi:2015zna,Bah:2017wxp} building from earlier work in \cite{Maldacena:2000mw}.\footnote{Holographic duals of twisted compactifications of five-dimensional SCFTs were also studied in \cite{Karndumri:2015eta} using six-dimensional $F(4)$ supergravity.  There, twist of a $U(1)$ flavor symmetry was considered by adding a vector multiplet to the theory.  However it was not proven that the theory considered is a consistent truncation of massive IIA supergravity.}   

In section \ref{fivescft} we review the construction of the Seiberg theories and their $AdS_6\times S^4_{\frac{1}{2}}$ gravity duals obtained in \cite{Brandhuber:1999np}.  The half sphere, $S^4_{\frac{1}{2}}$, is due to the presence of a O8/D8 system.  We then discuss the possible topological twist of the five-dimensional theories on a Riemann surface that can preserve at least four supercharges.  In addition to the R-symmetry, we also discuss turning on background gauge fields of the flavor symmetry on the Riemann surface, referred to as flavor twists.  The $SU_\mathcal{M}(2)$ mesonic symmetry of Seiberg theories has a geometric origin in the construction of the SCFTs and therefore its twist can described by curvature in geometry.   

In section \ref{gravitydual} we argue that the quantum field theories, in the low-energy limit of the compactified Seiberg SCFTs, also describe the world-volume dynamics of $N$ D4-branes wrapped on the Riemann surface in an O8/D8 background.  From this perspective, supersymmetry allows us to identify the Riemann surface as a holomorphic curve of a Calabi-Yau three-fold that is a sum of two line bundles over the curve.  The phases of the line bundles can be identified with a $U(1)^2$ subgroup of the $SU_\mathcal{R}(2)$ R-symmetry and the $SU_\mathcal{M}(2)$ mesonic symmetry of the five-dimensional SCFTs.  The degrees of the line bundles are then the twist parameters for the Cartan $U(1)^2$ subgroup of $SU_\mathcal{R}(2) \times SU_\mathcal{M}(2)$.  From the D4-branes picture we are able to construct the most general $AdS_4$ solutions that can be dual to three-dimensional SCFTs in the low-energy limit of the five-dimensional SCFTs.  The system derived is exactly analogous to the general holographic duals of the four-dimensional SCFTs on the world-volume of M5-branes on a punctured Riemann surface \cite{Bah:2015fwa}.   

In section \ref{constantc} we construct explicit $AdS_4\times \Sigma_g \times \widetilde{S}^4_{\frac{1}{2}}$ solutions dual to cases when the Riemann surface is smooth and without punctures.  The $\widetilde{S}^4_{\frac{1}{2}}$ is a deformed half-sphere that descends from the parent half-sphere in the $AdS_6 \times S^4_{\frac{1}{2}}$ duals of Seiberg SCFTs.  The solutions are exactly analogous to the B$^3$W solutions that describe the twisted compactifications of M5-branes wrapping a smooth Riemann surface in M-theory \cite{Bah:2012dg}\footnote{These similarities are due to the fact that the Riemann surface in both cases is embedded in a $CY_3$ that is always a sum of two line bundles. This is related to the universality discussed in \cite{Bobev:2017uzs}.} except for an additional contribution to the overall warp factor due to the O8/D8 system.  We compute some observables for the dual field theories from these solutions: the free energy and scaling dimensions of operators dual to D2-branes that wrap internal cycles.  The free energy, for example, is 
\begin{equation}\label{FEintro}
 \mathcal{F}=\frac{8\pi }{5 }\frac{ 2(1-g) N^{5/2}} {\kappa \sqrt{8-N_f}} F_z(z,\kappa), \qquad F_z(z,\kappa)=\frac{(|z^2-\kappa^2|)^{3/2}(\sqrt{\kappa^2 +8z^2} -\kappa)}{(|4z^2 -\kappa^2 + \kappa \sqrt{\kappa^2 + 8z^2}| )^{3/2}}
 \end{equation} where $g$ is the genus of the Riemann surface, $\kappa\in\{-1,0,1\}$ is its sectional curvature, and $z$ is the twist parameter for the $U(1)$ in $SU_\mathcal{M}(2)$.  The torus limit can be taken.  We observe the $\frac{N^{5/2}}{\sqrt{8-N_f}}$ scaling due to the five-dimensional origin of the SCFTs.\footnote{A natural name for these theories is: theories of class $\mathcal{F}$ due to their five-dimensional origin.   Class $\mathcal{S}$ theories are similarly named.  However the authors of \cite{Lawrie:2018jut} define theories of class $F$ where $F$ refers to fiber bundles.  We hope they can appreciate the symmetry in our naming.}

%%%%%%%%%%%%%%%%%%%%%%%%%%%%%%%%%%%%%%%%%%%
%%%%%%%%%%%%%%%%%%%%%%%%%%%%%%%%%%%%%%%%%%%
\section{Five-Dimensional SCFTs with Twist}\label{fivescft}

In this section we review the construction of the five-dimensional $\mathcal{N}=1$ SCFTs of interest, and then discuss the possible topological twists that can lead to interacting  three-dimensional $\mathcal{N}=2$ SCFTs.  

\subsection{5D SCFTs from D4-O8/D8}

We briefly review the construction of five-dimensional SCFTs from $N$ D4-branes probing a type IIA background geometry with an O8-plane and $N_f$ D8-branes \cite{Seiberg:1996bd}.  We also review their holographic duals constructed in \cite{Brandhuber:1999np} (see \cite{Bergman:2012kr} for a review and generalizations with branes at orbifolds).  The space-time is split as $\mathbb{R}^{1,4} \times \mathbb{R}^4 \times \mathbb{R}$ and the branes are distributed as:

\begin{center}
\begin{tabular}{c | c c c c c c c c c c}
& $x_0$ & $x_1$ & $x_2$ & $x_3$ & $x_4$ & $x_5$ & $x_6$ & $x_7$ & $x_8$ & $x_9$ 
\\
\hline 
D4 & $\times$ & $\times$ &  $\times$ &  $\times$ &  $\times$ & -- &  -- & -- & -- & -- \\
O8/D8 & $\times$ & $\times$ &  $\times$ &  $\times$ &  $\times$ & $\times$ &  $\times$ & $\times$ & $\times$ & --
\end{tabular}
\end{center}

The D4-branes are localized along the $56789$ directions, while the O8/D8 are localized on $x_9$.  More precisely, the O8-plane sits at $x_9=0$ and $N_f$ D8-branes  are distributed along the $x_9$ directions, with the $i$'th brane sitting at $x_9=x_9^i$.  The $N$ D4-branes sit at $x_5=x_6=x_7=x_8=0$ for simplicity, and distributed along the $x_9$ direction.  
The background metric of the O8/D8 system in type IIA supergravity in the string frame is
\begin{align}
ds^2_{10} &= H_8^{-1/2} \left( -dx_0^2 + dx_1^2 \cdots dx_8^2 \right) + H_8^{1/2} dx_9^2, \qquad e^{\phi} =g_s H_8^{-5/4} \\
H_8  &= c_0 + \frac{8 g_s}{2\pi \ell_s} x_9 -  \sum_i^{N_f} \frac{g_s |x_9 -x_9^i|}{4\pi \ell_s} -\sum_i^{N_f} \frac{g_s |x_9 +x_9^i|}{4\pi \ell_s}.
\end{align}  The background solution also admits a piece-wise constant Romans mass given by the $F_0$ RR-form:
\begin{equation}
F_0 =\frac{1}{2\pi}  \begin{cases} 8 \qquad &\mbox{for} \qquad x_9 < x_9^1 \\ 8-i \qquad &\mbox{for} \qquad x_9^i < x_9 < x_9^{i+1}.  \end{cases} 
\end{equation}
Our primary interest is on the world-volume theory of the probe D4-branes.  In the case of a single probe brane that is not coincident with the O8-plane, the world-volume theory is a five-dimensional $U(1)$ gauge theory with $\mathcal{N}=1$ supersymmetry.  The scalar in the vector multiplet parametrize the fluctuations of the brane along $x_9$.  There are $N_f$ hypermultiplets corresponding to open strings between the D4-brane and the $N_f$ D8-branes, with masses given by the relative positions of the branes -- refer to them as $H_f$.  There is also a hypermultiplet (with four real scalars) that parametrizes the deformations of the D4-brane along the $5678$ directions -- refer to it as $H_t$.

In the case where the brane is coincident with the O8-plane, the gauge group enhances to $USp(2) \cong SU(2)$.  When there are $N$ branes coincident with the O8-plane, the gauge group enhances to $USp(2N)$.  The $H_f$ hypermultiplets are in the fundamental representation of the gauge group while the $H_t$ hypermultiplet is in an antisymmetric representation.  

The world-volume theory can admit a large global symmetry in addition to supersymmetry.  When the D8-branes are all coincident, there is a $SU(N_f)$ global symmetry when they are off the O8-plane,  and it further enhances to $SO(2N_f)$ when the D8-branes are coincident with the O8-plane.  The $H_f$ hypermultiplets transform in the fundamental representation while the $H_t$ hypermultiplet is neutral.  The world-volume theory admits an $SO(4)\cong SU_{\mathcal{R}}(2)\times SU_{\mathcal{M}}(2)$ symmetry corresponding to the rotations of the $5678$ directions.  The $SU_\mathcal{R}(2)$ is the R-symmetry associated to the $\mathcal{N}=1$ five-dimensional supersymmetry, with the two pseudoreal supercharges transforming as a doublet.  The $SU_\mathcal{M}(2)$ part is a flavor symmetry under which the four real scalars in the $H_t$ hypermultiplet transform as a complex doublet.  

Given a five-dimensional gauge symmetry of field strength, $F$, the $U(1)$ current $j = *_5 \mbox{Tr}(F\wedge F)$ is conserved.  This implies a $U_I(1)$ flavor symmetry which corresponds to conservation of instanton number for the five-dimensional gauge theory.  From the IIA set-up, this $U_I(1)$ corresponds to RR $C_1$ potential restricted along the world-volume of the D4-branes.  The total global symmetry of the world-volume theory of the probes when all branes are coincident at $x_9=0$ is $SU_\mathcal{R} (2) \times SU_\mathcal{M}(2) \times SO(2N_f) \times U_I(1)$.  

The effective gauge coupling of the world-volume theory can be computed by expanding the DBI action of the D4-branes, and is given as
\begin{equation}
\frac{1}{g_{YM}^2} = \frac{1}{g_s} \frac{2 H_8}{2\pi \ell_s} = \frac{1}{g^2_{cl}} + \frac{16 x_9^0}{(2\pi \ell_s)^2} - \sum_i^{N_f} \frac{ |x_9^0 -x_9^i|}{(2\pi \ell_s)^2} -\sum_i^{N_f} \frac{ |x_9^0 +x_9^i|}{(2\pi \ell_s)^2},
\end{equation} where $x_9^0$ is the location of the D4-branes and $g_{cl}^2=\frac{g_s c_0}{2\pi \ell_s}$ is the classical coupling constant.  The field theory limit is obtained by taking $\ell_s \to 0$.  Notice that in this limit, $g_{cl}$ blows up, and the primary question here is whether there can be a fixed point where $g_{YM}$ is finite.  To this end, it is natural to define $\phi = \frac{x_9^0}{(2\pi \ell_s)^2}$ as the parameter for the Coulomb branch of the world-volume gauge theory.  This parameter is necessarily positive, thereby identifying the Coulomb branch with $\mathbb{R}^+$.  The field theory limit to a point on the Coulomb branch is
\begin{equation}
g_{YM}^2 = \mbox{fixed}, \qquad \phi = \mbox{fixed}, \qquad \ell_s \to 0.  
\end{equation}  This limit necessarily puts the D4-branes on the O8-plane where there is a $USp(2N)$ gauge symmetry.  The effective gauge coupling is 
\begin{equation}
\frac{1}{g^2_{YM}} = \frac{1}{g^2} + 16 \phi - \sum_i^{N_f} |\phi - m_i| - \sum_i^{N_f} |\phi + m_i|
\end{equation} where the mass parameters $m_i = \frac{x_9^i}{(2\pi \ell_s)^2}$ of the $H_f$ hypermultiplets parametrize the Higgs branch of the theory.  The Coulomb branch can be studied by going to the origin of the Higgs branch ($m_i=0$) where there is a $SO(2N_f)$ flavor symmetry.  The gauge coupling in the Coulomb branch is finite in the decoupling limit only when 
\begin{equation}
n_0 \equiv 8-N_f >0.  
\end{equation} Therefore one can expect a strongly-coupled fixed point at the origin of the moduli space when $N_f \leq 7$.  In \cite{Seiberg:1996bd} Seiberg argued that there does exist strongly coupled five-dimensional SCFTs in the decoupling limit.  Moreover the $SO(2N_f) \times U_I(1)$ flavor symmetry enhances to an exceptional group, $E_{N_f+1}$.  There are eight of such SCFTs with a global symmetry $SU_\mathcal{R} (2) \times SU_\mathcal{M}(2) \times E_{N_f +1}$.  

One can consider the system beyond the probe limit, where the $N$ D4-branes backreact.  In this case one can look for an $AdS_6$ near-horizon geometry dual to the the SCFTs.  This is the celebrated result by Brandhuber and Oz in \cite{Brandhuber:1999np}.  The holographic dual is given by the geometry
\begin{equation}
ds^2_{10}= \ell_s^2 \, \Omega^{1/2} \left[ds^2(AdS_6) + \frac{4}{9}\left(\frac{d\mu_0^2}{1-\mu_0^2} +  (1-\mu_0^2)ds^2(S^3) \right) \right], \qquad \Omega = \frac{18\pi^2 N}{n_0\mu_0^{2/3}},
\end{equation}
where $ds^2(AdS_6)$ is the line element of the unit-radius $AdS_6$ and $ds^2(S^3)$ the line element of the unit-radius three-sphere. The range of $\mu_0$ is $(0,1]$, implying that the internal four-manifold is a half four sphere, $S_{1/2}^4$. The dilaton and the four-form flux are
\begin{equation}
e^{-4\phi} = \frac{9N n_0^3 \mu_0^{10/3}}{8 \pi^2}, \qquad F_4 = \frac{80}{9} \ell_s^3 \pi N (1-\mu_0^2) \mu_0^{1/3} d\mu_0 \wedge {\rm vol}(S^3) .
\end{equation}

\subsection{Topological Twist of 5D SCFTs}

In this section, we consider the compactification of the five-dimensional SCFTs on a two-dimensional Riemann surface with genus, $g$, and $n$ punctures, $\Sigma_{g,n}$.  Such compactification generically breaks supersymmetry; however a number of supercharges can be preserved by a partial topological twist \cite{Witten:1988ze,Bershadsky:1995qy}.  This is a prescription for obtaining constant and globally defined background spinors that can be identified with supercharges.  
The Killing spinor equation in the presence of R-symmetry is 
\begin{equation} 
\left(\partial_\mu + \frac{1}{4} \omega_\mu + A_\mu \right) \varepsilon =0,
\end{equation} where $\omega_\mu$ is the spin connection and $A_\mu$ is a background gauge field valued in the R-symmetry.  A constant spinor is obtained by turning on a background field to cancel the spin connection.  In effect this identifies the holonomy group of the curved geometry, in which $\omega_\mu$ is valued, with a subgroup of the R-symmetry.  In the case of interest, the Riemann surface has a $U_h(1)$ holonomy group and its generator, $J_h$, is identified with the generator of the $U_\mathcal{R}(1)$ subgroup of $SU_\mathcal{R}(2)$, $J_\mathcal{R}^3$.  Under the twist, only the $U_\mathcal{R}(1)$ part of the R-symmetry is preserved by the compactification.  Generically this twist will preserve four supercharges and therefore $\mathcal{N}=2$ in three dimensions. 

The five-dimensional theory admits a larger global symmetry than the R-symmetry.  In addition to $J_\mathcal{R}^3$, the generator of the holonomy group can be identified with a $U(1)$ subgroup of the global symmetry.  So we can consider the $U(1)$ generators of the flavor symmetry as $J_\mathcal{M}^3$ and $J_f^i$ for the Cartan elements of $SU_\mathcal{M}(2)$ and $E_{N_f+1}$ respectively.  Under the compactification, we can generally write
\begin{equation}
J_h =  J_{\mathcal{R}}^3 + z J_{\mathcal{M}}^3 + z_i J_f^i.  
\end{equation} The coefficient of $J_{\mathcal{R}}^3$ is fixed by the topological twist.  Unlike the coefficients of $J_\mathcal{R}^3$, the other twist parameters $(z,z_i)$ are not constrained by the topological twist.  Another way to understand these more general embeddings of the holonomy group is to consider the background gauge fields valued in the global symmetry.  As already discussing in the cases of the R-symmetry, the twist reflects the presence of a non-trivial background gauge field that has legs along the Riemann surface.  More precisely, if we consider the five-dimensional background gauge fields for the Cartan $U(1)$'s $(A_{\mathcal{R}}, A_{\mathcal{M}}, A^f_i)$, these are related to the three-dimensional background gauge fields $(a_{\mathcal{R}}, a_{\mathcal{M}}, a^f_i)$ as
\begin{equation}
A_{\mathcal{R}} = a_\mathcal{R} - V, \qquad  A_{\mathcal{M}} = a_{\mathcal{M}} - z V, \qquad A^{f}_i = a^{f}_i - z_i V.
\end{equation} The one-form $V$ is the spin connection of the surface and satisfies
\begin{equation}
\frac{1}{2\pi} \int_{\Sigma_{g,n}} dV = 2(1-g) -n \equiv \chi.  
\end{equation}  The twist parameters $(z,z_i)$ are fluxes over the Riemann surface and are quantized as
\begin{equation}
\frac{1}{2\pi} \int_{\Sigma_{g,n}} dA_{\mathcal{M}} = -z \chi \in \mathbb{Z}, \qquad \frac{1}{2\pi} \int_{\Sigma_{g,n}} dA^{f}_i = -z_i \chi \in \mathbb{Z}.  
\end{equation}  

In the special case of the torus, where the Euler characteristic $\chi$ vanishes, a non-trivial topological twist is not needed to preserve supersymmetry, and therefore the component of $A_{\mathcal{R}}$ along the surface vanishes.  However, we can still turn on non-trivial twist parameters for $A_\mathcal{M}$ and $A^{f}_i$.  In such case, one picks a one-form $V$ on the torus  that satisfies
\begin{equation}
\int_{T^2} dV = 1.
\end{equation} The quantization of the $(z,z_i)$ follows.  

If only the R-symmetry is twisted, the resulting three-dimensional theory admits the full flavor symmetry $SU_\mathcal{M} (2)\times E_{N_f+1}$ and a $U_\mathcal{R}(1)$ R-symmetry.  However if we consider a more generic twist, then the global symmetry of the low-energy theory is $U_\mathcal{R}(1) \times U_\mathcal{M}(1) \times U(1)^r$ where $r$ is the rank of $E_{N_f+1}$.  Depending on the choice of twist parameters, a larger subgroup of $SU_\mathcal{M} (2)\times E_{N_f+1}$ can be preserved.  

In this paper we will worry primarily about twisting geometric symmetries, i.e. the $U(1)$ subgroups of the $SO(4)$ symmetry, which has its origin from the isometry of the directions transverse to the D4-branes.  The other symmetries have their origin from the D8-branes and from the RR $C_1$ potential.  The $U_\mathcal{R}(1) \times U_\mathcal{M}(1)$ corresponds to the Cartan elements of the $SO(4)$ symmetry, and we can make a democratic choice of the generators as
\begin{equation}
J^+ = J_\mathcal{R}^3 + J_\mathcal{M}^3, \qquad J^- = J_\mathcal{R}^3 - J_\mathcal{M}^3.  
\end{equation} In this picture the holonomy group is embedded as
\begin{equation}
J_h = \frac{p}{p+q} J^+ + \frac{q}{p+q} J^- + z_i J_f^i, \qquad p = - \frac{1+z}{2} \chi, \quad q = - \frac{1-z}{2} \chi.  \label{twist}
\end{equation}

If the three-dimensional low energy theory is a superconformal field theory, then its superconfomal R-symmetry is fixed by $\mathcal{F}$-extremization \cite{Jafferis:2010un}.  The generator of the trial R-symmetry is given as
\begin{equation}
J_R = a_+ J^+ + a_- J^- + s_i z_i J_f^i, \qquad a_{\pm} = \frac{1\pm \epsilon}{2}.
\end{equation}  The parameters $(\epsilon, s_i)$ are fixed by $\mathcal{F}$-extremization.  Only the twisted generators from $E_{N_f+1}$ can mix with the R-symmetry, and we enforce this by explicitly including $z_i$ in the trial R-symmetry.  

In summary the bosonic symmetries of the five-dimensional quantum field theory from the bulk space-time are broken as
\begin{equation}
SO(2,4) \times SO(4) \ \Longrightarrow \ SO(1,2) \times U_+(1) \times U_-(1) \label{symb}
\end{equation} where $U_{\pm}(1)$ are the symmetries with generators $J^\pm$.  Our interest now is to argue that the three-dimensional field theories are SCFTs and therefore the $SO(1,2)$ enhances to the superconformal group, $SO(2,3)$.  

%%%%%%%%%%%%%%%%%%%%%%%%%%%%%%%%%%%%%%%%%%%%%%%%%%%%%%%%%%%%%%%%%
%%%%%%%%%%%%%%%%%%%%%%%%%%%%%%%%%%%%%%%%%%%%%%%%%%%%%%%%%%%%%%%%%%%

\section{Gravity Dual}\label{gravitydual}

In this section we discuss how to construct the holographic duals to the three-dimensional SCFTs in the low-energy limit of five-dimensional SCFTs compactified on a Riemann surface, $\Sigma_{g,n}$.  First, we discuss the general ansatz for such $AdS_4$ solutions in type IIA supergravity and then describe the general system dual to the SCFTs.  In appendix \ref{derivation} we show in detail how to obtain the solutions from the system recently derived in \cite{Passias:2018zlm}.  

\subsection{Ansatz}

Our goal is to study the compactification of the five-dimensional SCFTs from a top-down perspective.  We will construct the $AdS_4$ duals to the three-dimensional SCFTs by thinking about the original D4-branes probing the O8/D8 background.  The main expectation is that the low energy limit of the five-dimensional SCFTs on $\Sigma_{g,n}$ describes also the low-energy limit of the world-volume theory of the D4-branes wrapped on $\Sigma_{g,n}$ in the O8/D8 background.  The discussion will be exactly anologous to the compactification of M5-branes on a Riemann surface in M-theory as described in section 2 of \cite{Bah:2012dg} and section 3 of \cite{Bah:2015fwa}.

First write the O8/D8 background as
\begin{equation}
ds^2_{10} = H_8^{-1/2} ds^2_9\left(M^{1,8} \right) + H^{1/2} ds^2_9.  
\end{equation}  Now we decompose $M^{1,8}$ as
\begin{equation}
M^{1,8} \; \rightarrow \; \mathbb{R}^{1,2} \times CY_3.
\end{equation}  The $N$ D4-branes are extended along $\mathbb{R}^{1,2}$ and along a holomorphic curve $\mathcal{C}_{g,n}$ embedded in the $CY_3$.  In the region near the brane, the $CY_3$ is a $U(2)$-bundle over the $\mathcal{C}_{g,n}$ whose determinant line bundle is fixed to the the canonical bundle of the Riemann surface.  In this paper, we restrict to the case where the structure group is $U(1)^2$ (see section 2 of \cite{Bah:2012dg} for more details).  The $CY_3$ in this case is a sum of two line bundles $\mathcal{L}_{p} \oplus \mathcal{L}_{q}$ over the Riemann surface of degree $p$ and $q$.  The local geometry is given by
\begin{equation}
\begin{array}{ccc}
\mathbb{C}^2 & \hookrightarrow & \mathcal{L}_{p} \oplus \mathcal{L}_{q} \\ & & \downarrow \\ & & \mathcal{C}_{g,n}.
\end{array}
\end{equation} The vanishing of the first chern class of the $CY_3$ implies the twist condition $p+q= -\chi$ as in \eqref{twist}.  The phases of the line bundles correspond to the symmetries $U_+(1)\times U_-(1)$ in  \eqref{symb} from the field theory perspective.  

The goal is to check whether the low-energy limits of world-volume theory of the D4-brane configuration in the O8/D8 background are captured by three-dimensional SCFTs.  This problem can be studied in holography by looking for $AdS_4 \times_w M_6$ solutions with the appropriate symmetries in massive type IIA supergravity; the local metric has the form
\begin{equation}
ds^2_{10} = e^{2A} \left[ ds^2(AdS_4) + ds^2(M_6) \right] \label{genmet}
\end{equation} for some warp factor $e^{2A}$.  The internal manifold $M_6$ is constrained as 
\begin{equation}
\begin{array}{ccc}
M_4 & \hookrightarrow & M_6 \\ & & \downarrow \\ & & \Sigma_{g,n}.
\end{array}
\end{equation} where $\Sigma_{g,n}$ is IR limit of the curve $\mathcal{C}_{g,n}$.  The four-manifold, $M_4$, admits at least a $U(1)^2$ isometry group corresponding to the phases of the line bundles, $U_+(1) \times U_-(1)$.  These symmetries also correspond to the structure group of $M_4$ over the $\Sigma_{g,n}$.  The ten-dimensional space-time has the form
\begin{equation}
M^{1,9} \quad \to \quad AdS_4 \times_w\left( \Sigma_{g,n} \times S^1_+ \times S_-^1 \times [t^+] \times [t^-] \right) \label{ansatz}
\end{equation} where the isometries from the circles, $S_\pm^1$, are dual to $U_\pm(1)$.  The last two directions with coordinates $(t^{+},t^-)$ do not, generically, correspond to any isometries.  The $AdS_4$ radius and the intervals $(t^{+},t^{-})$ are related to the radius of the line bundles and to $x_9$.  The most general ansatz has many component functions and depends on the Riemann surface coordinates and the $t$ coordinates.  However the system is nevertheless tractable in supergravity.  

Recently, the authors of \cite{Passias:2018zlm} presented the conditions for $\mathcal{N}=2$ supersymmetric $AdS_4$ solutions in massive type IIA supergravity, and in particular they provided the explicit system for the case that $M_6$ is a manifold with $SU(2)$-structure.  We can look for the $AdS_4$ solutions of interest within this class.  The general form of the metric is 
\begin{equation}
ds^2_{10} =e^{2A} \left[ ds^2(AdS_4) + \frac{dy^2}{e^{4A}-y^2} + \frac{e^{4A} -y^2}{4e^{4A}} \left(d\psi + \rho\right)^2 + \frac{y}{\ell^2 + y^2 F_0 e^{4A}} ds^2(N_4) \right]. \label{ogmetric}
\end{equation} The warp factor, $e^{2A}$, and the metric depend on $y$ and on the coordinates on the four-manifold, $N_4$.  At fixed $y$ the space $N_4$ is K\"{a}hler.  The supergravity equations are given in terms of the K\"ahler and holomorphic two-forms of $N_4$.  The $\psi$ direction parametrizes a circle fibered over $N_4$; its connection, $\rho$, is completely fixed by the base four-manifold.  The $U(1)$ generated by $\partial_\psi$ is dual to the superconformal R-symmetry.    

For the problem of interest, we make the ansatz
\begin{equation}
ds^2 (N_4) = e^{2W}\left(dx_1^2 + dx_2^2\right) + e^{2Z} \left[\left(d\tau+V^R \right)^2 + e^{2C} \left(d\phi+ V^I\right)^2 \right].  
\end{equation} The $x_i$ coordinates parametrize the Riemann surface directions.  We fix $\ell=0$ to restrict to solutions with D4-branes.  In addition to the R-symmetry circle $\psi$, we also demand a circle on the base manifold in order to obtain the desired $U(1)^2$ isometry.  The one forms $V^{I/R}$ have ``legs'' on the Riemann surface.  All metric functions depend on $x_i$, $\tau$ and $y$.  In appendix \ref{derivation} we describe the the reduction of the supersymmetry equations and obtain the general form of the desired internal manifold. 

We can identify the isometry $\partial_\phi$ with the flavor symmetry in the dual theory.  From this we obtain
\begin{equation}
\partial_\phi = \frac{1}{2} \left(\partial_{\phi_+} - \partial_{\phi_-}\right), \qquad \partial_\psi = -a_+ \partial_{\phi_+} - a_- \partial_{\phi_-}, \qquad J^\pm \cong \partial_{\phi_\pm}.
\end{equation} These identifications allow us to reduce the supersymmetry system to a more useful description of the gravity duals.  We describe them next.

\subsection{General $AdS_4$ System}

The gravity duals to the three-dimensional $\mathcal{N}=2$ SCFTs that describe the low-energy limit the D4-branes system are given by
\begin{equation}
ds^2_{10} = e^{2A}\left[ ds^2(AdS_4) + \frac{1}{4} H \left( e^{2w} \left(dx_1^2 + dx_2^2\right) + 4 h^{ij} \eta_i \eta_j + g_{ij} dt^i dt^j \right) \right],  \label{pmmetric}
\end{equation} where $(i,j)$ take $\pm$ values.  The components of the metric are given in terms of a single potential, $D_0$, as
\begin{equation}
g_{ij} = - \partial_{i} \partial_j D_0, \qquad h_{ij} = - \partial_{i} \partial_j \widetilde{D}_0, \qquad \widetilde{D}_0 = D_0 + \frac{4}{3} t \left(\ln t -1 \right), \quad h_{ij} h^{jk} = \delta^k_i
\end{equation} where $\det(h)$ and $\det(g)$ are the determinants of $h_{ij}$ and $g_{ij}$ respectively.  The warp factors are given as
\begin{equation}
e^{2A} =\frac{H^{-1/2}}{\sqrt{y} F_0}, \qquad H = \frac{1}{3t} \frac{\det(h)}{\det(g)}, \qquad e^{2w} = 3t \det(g) e^{(\partial_+ + \partial_-) \widetilde{D}_0}.  
\end{equation} The one-forms and the $y$ function are given as
\begin{equation}
\eta_\pm = d\phi_\pm + \frac{1}{2} \star_2 d_2 \partial_\pm D_0, \qquad y^3 F_0^2 = 3t = 3(a_+ t^+ + a_- t^-)  
\end{equation} where $\star_2$ and $d_2$ are taken over the Riemann surface directions with coordinates $(x_1,x_2)$.  Our conventions for the Hodge star is $\star_2 dx_1 = -dx_2$.  

The full system is governed by a single potential, $D_0$, that satisfies
\begin{equation}
\left(\partial_{x_1}^2 + \partial_{x_2}^2 \right) D_0 = e^{2w}.  
\end{equation}  The solutions are supported by $F_0$, due to the O8/D8 system, a running dilaton and a four-form flux. The flux is most naturally written in terms of the variables
\begin{equation}
t = a_+ t^+ + a_- t^-, \qquad u = t^+-t^-, \qquad \eta_t = \eta_+ + \eta_-, \qquad \eta_u = a_- \eta_+ - a_+ \eta_-. \label{tucoord}
\end{equation} 
The dilaton and four-form flux then take the form
\begin{equation}
e^{-4\phi} = y^5F_0^6 H, \qquad F_0 F_4 = \left[ J -\frac{1}{2} dt \wedge \left(\eta_t -A_u \eta_u \right) \right] \wedge K - \frac{1}{2y} J\wedge J, \label{flux}
\end{equation} where $J$, $K$, and $A_u$ are given by
\begin{align}
J &= \frac{1}{4} e^{2w} dx_1 \wedge dx_2 - \frac{1}{2} \left(du + A_u dt \right) \wedge \eta_u, \qquad A_u = \frac{\partial_t \partial_u D_0}{\partial_u^2 D_0} \label{Jdef}\\
K &= \frac{1}{2y} \left(1- 3t H \right) d \left[ \eta_t - A_u \eta_u\right]-\frac{1}{2} d \left(\frac{3t}{y} H\right) \wedge \left(\eta_t -A_u \eta_u \right).
\end{align}
The circle one-forms $(\eta_t-A_u \eta_u, \eta_u)$ are respectively dual to the superconformal R-symmetry and to the Cartan $U(1)$ of $SU_\mathcal{M} (2)$.  When the connection form of circle dual to $\eta_u$ vanishes, its associated $U(1)$ isometry can enhance to the full $SU_\mathcal{M}(2)$.  

There is another enhancement of the symmetry of the internal metric.  When the connection form of the circle dual to $\eta_+$ or $\eta_-$ vanishes, the corresponding $U(1)$ isometry can enhance to an $SU(2)$ isometry.\footnote{This is also true when the connection form of a circle dual to any linear combination of $\eta_\pm$ vanishes.  This follows from a symmetry of the metric that simultanously rotates the circle directions and the $t$ coordinates,  See section 3.2 of \cite{Bah:2015fwa}.}  In this case, the internal metric admits twice as many Killing spinors.  However this enhanced supersymmetry is broken by the O8/D8-branes since the overall $\sqrt{y}$ in the warp factor will depend on coordinates of the internal two-sphere associated to the enhanced $SU(2)$ isometry.  From the probe D4-branes perspective, this corresponds to the case when $CY_3 \cong CY_2 \times \mathbb{C}$, where $CY_2$ is the cotangent bundle of the Riemann surface.  This is a configuration where the D4-branes preserve more supersymmetry which is however broken down by the presence of the O8/D8-system.

\section{Constant Curvature Ansatz}\label{constantc}

In this section we restrict the general $AdS_4$ system above to cases when the Riemann surface has constant sectional curvature.  This restriction can be made without loss of generality whenever the Riemann surface is smooth and admits no punctures.  From this point of view the seemingly complicated system exists to support solutions where the Riemann surface has punctures, as in \cite{Gaiotto:2009gz,Bah:2015fwa}.  

The ansatz for the conformal factor of the $(x_1,x_2)$ plane is
\begin{align}
e^{2w}=f(t^+,t^-)e^{2A_0(x_1,x_2)},\qquad \left(\partial_{x_1}^2 + \partial_{x_2}^2 \right)A_0=-\kappa e^{2A_0} \label{warpsep}
\end{align}
where $\kappa$ is the curvature of the surface; it takes value from $\{-1,0,1\}$ for $H_2$, $T^2$ and $S^2$, respectively.  We can replace $H_2$ with $H_2/\Gamma$ to obtain a constant curvature Riemann surface with genus, $g$; $\Gamma$ is an element of the Fuschian subgroup of the $PSL(2,R)$ isometry group of $H_2$.  The function $f$ is to be determined.  A representative solution for $A_0$ can be written as
\begin{equation}
e^{A_0} = \frac{2}{1+ \kappa \left(x_1^2 + x_2^2\right)}.  
\end{equation}  It is convenient to introduce the connection one-forms $V$ defined as
\begin{equation}
V =  \alpha \frac{x_1 dx_2 - x_2 dx_1}{1+ \kappa \left(x_1^2 + x_2^2\right)}, \qquad \alpha = \begin{cases} &\frac{1}{4 \pi} \quad \mbox{for} \quad \kappa=0 \\ &\frac{\kappa}{1-g} \quad \mbox{for} \quad \kappa \neq 0 \end{cases}.  \label{Vandal}
\end{equation} These can also be written as
\begin{equation}
V=-\frac{1}{2} \alpha \star_2 d_2 \widetilde{A}^0, \qquad \widetilde{A}_0 = \begin{cases} - \left(x_1^2 + x_2^2\right) \quad &\mbox{for} \quad \kappa=0 \\
\kappa A_0 \quad &\mbox{for} \quad \kappa \neq 0  \end{cases}.
\end{equation} The normalizations are such that 
\begin{equation}
\left(\partial_{x_1}^2 + \partial_{x_2}^2 \right)\widetilde{A}_0 = - e^{2A_0}, \qquad dV = \frac{\alpha}{2} e^{2A_0} dx_1 \wedge dx_2, \qquad \int dV = 2\pi.  
\end{equation}

The separability condition in \eqref{warpsep} implies that $D_0$ has the form
\begin{equation}
D_0 =- \mu_g^2 \widetilde{A}_0 + I_0(t^+,t^-), \qquad \mu_g^2 = c_0 - 2m_+ t^+ - 2m_- t^-. 
\end{equation} The parameters must satisfy
\begin{equation}
m_+ + m_- = \kappa, \qquad m_\pm = \frac{\kappa \pm z}{2}, 
\end{equation} where $z$ is a free parameter that is related to the twist parameter for the flavor $U_\mathcal{M}(1)$.  The one-forms dual to the isometry generators are
\begin{equation}
\eta_\pm = d\phi_\pm - \frac{2m_\pm}{\alpha} V.  
\end{equation}  Plugging the ansatz for $D_0$ into the equation of motion yields a Monge-Amp\'{e}re equation for $I_0$ given as
\begin{equation}
3 t^{7/3} \left[\partial_+^2 I_0 \partial_-^2 I_0 - (\partial_+ \partial_- I_0)^2 \right] e^{(\partial_+ + \partial_-)I_0} = \mu_g^2.  
\end{equation}

\subsection{$(p,q)$ Solutions}
Now we consider the class of solutions given by the ansatz\footnote{There is a larger class of ans\"{a}tze that we can consider.  See section 4 of \cite{Bah:2015fwa}.}  
\begin{equation}
I_0 = - \frac{4}{3} t (\ln t -1) - \frac{1}{2}\mu_1^2 (\ln \mu_1^2 -1) - \frac{1}{2}\mu_2^2 (\ln \mu_2^2 -1) + 2(t^+ + t^-) \nu,
\end{equation} where it is convenient to define new coordinates as
\begin{equation}
\mu_0^2 = 2 t, \qquad \mu_1^2 = 1 -2 t^+, \qquad \mu_2^2 = 1-2 t^-, \qquad \mu_0^2 + a_+ \mu_1^2 + a_-\mu_2^2 = 1.  \label{mucord} 
\end{equation} The equation for $I_0$ reduces to an algebraic condition: 
\begin{equation}
\mu_g^2 = 2e^{2\nu} \left(3\mu_0^2 + 4 (a_+^2 \mu_1^2 + a_-^2 \mu_2^2 ) \right).
\end{equation} The solution to the algebraic relation is 
\begin{equation}
e^{2\nu} =\frac{ - \kappa \pm \sqrt{\kappa^2 + 8z^2}}{4}, \qquad \epsilon = \frac{\kappa \pm \sqrt{\kappa^2 + 8z^2}}{4z}, \qquad a_\pm = \frac{1\pm \epsilon}{2}.  \label{nuep}
\end{equation}  The metric of the full system is given by
\begin{align}
ds^2_{10} &= \frac{H^{-1/2}}{\sqrt{y} F_0} \left[ ds^2(AdS_4) + e^{2\nu} ds^2(\Sigma_g) + \frac{1}{4} H ds^2(M_4) \right] \\ \label{pqsol}
ds^2(M_4) &= \frac{8}{3} d\mu_0^2 + 2 \left( d\mu_1^2 + \mu_1^2 \eta_+^2 + d\mu_2^2 + \mu_2^2 \eta_-^2 \right).
\end{align} The warp factors are given by
\begin{equation}
H = \frac{2}{3\mu_0^2 + 4 (a_+^2 \mu_1^2 + a_-^2 \mu_2^2 )}, \qquad e^{2w} = \frac{4 e^{2\nu}}{H}, \qquad y^3 F_0^2 = \frac{3}{2} \mu_0^2.
\end{equation}  The manifold is compact when $a_\pm >0$, leading to the restriction $|\epsilon| <1$.  The positivity of the metric implies that we must take the $+$ solution in \eqref{nuep}.  The four-form flux is non-trivial, and can be written from the expression in \eqref{flux}.

In order to make the regularity conditions obvious and more easily study the geometry, we make the coordinate transformation
\begin{equation}
a_+ \mu_1^2 = (1-\mu_0^2)\cos^2(\theta), \quad a_- \mu_2^2 = (1-\mu_0^2) \sin^2(\theta),\quad q(\theta) = a_+ \cos^2(\theta) + a_- \sin^2(\theta),
\end{equation} to obtain the metric 
\begin{align}
ds^2(M_4) % &= \left(\frac{8}{3}+ \frac{2\mu_0^2}{q(\theta)(1-\mu_0^2) }\right) d\mu_0^2 + \frac{2(1-\mu_0^2)}{a_+ a_-} ds^2(M_3)\\
&=\frac{4}{3} \frac{1}{q(\theta) H}\frac{d\mu_0^2}{1-\mu_0^2} +\frac{8(1-\mu_0^2)}{1-\epsilon^2} ds^2(M_3)\\
ds^2(M_3) &= q(\theta)\left(d\theta - \epsilon \frac{\sin(2\theta)}{2q(\theta)} \frac{\mu_0 d\mu_0}{1-\mu_0^2 }\right)^2 +\frac{1-\epsilon}{2} \cos^2(\theta)\eta_+^2 + \frac{1+\epsilon}{2}\sin^2(\theta)  \eta_-^2.
\end{align}

The range of coordinates in $M_4$ are $\mu_0 \in [-1,1]$ and $\theta \in [0,\frac{\pi}{2}]$.  At the end-points of the $\theta$ angle, the $\phi_-$ and $\phi_+$ circles shrink.  The periodicity of the circles is $2\pi$, however the orientation is chosen so that $\phi_\pm$ takes value in $[0,-2\pi]$.  The quantization of the first Chern class of the $U(1)$ bundles over the Riemann surface imply
\begin{equation}
p \equiv -\frac{2m_+}{\alpha} \in \mathbb{Z}, \qquad q \equiv -\frac{2m_-}{\alpha} \in \mathbb{Z}, \qquad 2z= (q-p) \alpha, \qquad p+q = 2(g-1).  \label{pqpar}
\end{equation}  The parameter $\alpha$ depends on the $g$ and $\kappa$ as given in equation \eqref{Vandal}.   The integers $(p,q)$ are precisely the degree of the two line bundles of the Calabi-Yau three-fold. 

Individual solutions are labeled by the choice of two integers $(p,q)$ with the constraint in \eqref{pqpar}.  The metric shows this dependence through the genus $g$ and through $\epsilon$, which is given in \eqref{nuep} as a function of $z$ given a choice of $\kappa$.  The parameter $\epsilon$ takes discrete values since $z$ is quantized.  The bound on $\epsilon$ due to compactness of $M_4$ also implies a bound on $z$.  For $\kappa \in \{0,1\}$ we must have $z^2 >1$, and therefore $p$ and $q$ must have opposite signs. 

\subsubsection*{O8/D8 Region}

While $M_4$ is smooth, the full ten-dimensional metric is singular at $\mu_0=0$.  In this region there is a O8/D8 system.  This leads to the restriction
\begin{equation}
\mu_0 \in (0,1].
\end{equation}  The region near $\mu_0 =0$ can be written as
\begin{equation}
ds^2_{10} = \frac{\sqrt{q(\theta)}}{\sqrt{2y} F_0} \left[ds^2_{AdS_4} + e^{2\nu} ds^2(\Sigma_g) + \frac{1}{a_+ a_- q(\theta)} ds^2(M_3) \right] + \frac{\sqrt{2y} F_0}{\sqrt{q(\theta)}} dy^2 .
\end{equation}  with a dilaton
\begin{equation}
e^{-4\phi} =\frac{2y^5 F_0^6}{q(\theta)}.
\end{equation}  This metric does indeed describe the geometry of a coincident O8/D8 system with a warped $\left(AdS_4 \times \Sigma_g \right) \times_w M_3$ world-volume.

\subsubsection*{Flux Quantization}
The flux can be expressed as
\begin{equation}
 F_4 = \frac{1}{4} e^{2w} dx_1 \wedge dx_2 \wedge F_2 + F_4^\perp.  
\end{equation}  We need the $F_4^\perp$ term to compute the quantized flux as 
\begin{equation}
\frac{1}{(2\pi \ell_s)^3}\int F_4^\perp =N \in \mathbb{Z}.
\end{equation}  The parameter $N$ counts the number of D4-branes in the field theory construction.  The desired expression can be written as
\begin{align}
 F_4^\perp %&= \sin (2 \theta) \frac{ \mu_0^{1/3} (1-\mu_0^2)}{4F_0a_+ a_- y_0 \mu_g^4} \left[9\mu_0^2+8(1-\mu_0^2)(2a_+a_-+q(\theta))\right] d\mu_0  \wedge d\theta \wedge \eta_+ \wedge \eta_- \\
&= \sin (2 \theta) \frac{ \mu_0^{1/3} (1-\mu_0^2) H }{8a_+ a_- y_0 F_0} \left[3 + 2(1-\mu_0^2) \left(4a_+ a_- -q(\theta) \right) H \right] d\mu_0  \wedge d\theta \wedge \eta_+ \wedge \eta_-
\end{align}
 where $y_0= \left(\frac{3}{2F_0^2}\right)^{1/3}$. After integrating the expression, we obtain the quantization conditions of the flux, from which we obtain
 \begin{equation}
(2\pi \ell_s)^{8/3}  =  \frac{(2\pi)^2}{2}\left(\frac{3}{2}\right)^{2/3} \frac{4z^2 -\kappa^2 + \kappa \sqrt{\kappa^2+8z^2}}{2 (z^2-\kappa^2)n_0^{1/3}N} , \qquad F_0 = \frac{n_0}{2\pi \ell_s}
 \end{equation} in units where $L_{AdS_4}=1$.  
\begin{comment} 
 The function $I(\epsilon)$ is given as the integral 
\begin{align}
I(\epsilon) &= \int_{0}^1 \frac{\mu_0^{1/3} (1-\mu_0^2) (3\mu^2_0+16a_+a_-(1-\mu^2_0))}{9\mu_0^4 +12\mu_0^2(1-\mu_0^2)+ 16 a_+ a_- (1-\mu_0^2)^2}d\mu_0 \\
&+\color{red} \int_0^1 \mu_0^{1/3} \ln \left(\frac{3\mu_0^2 + 4a_+(1-\mu_0^2)}{3\mu_0^2 + 4a_-(1-\mu_0^2)} \right)\frac{d\mu_0}{2(a_+ -a_-)}. 
%\\&+ \int_0^1 2\mu_0^{1/3} \ln \left(\frac{a_++(4a_--1)\mu_0^2}{a_-+(4a_+-1)\mu_0^2} \right)\frac{d\mu_0}{a_+ -a_-}.    
\end{align} The function $I(\epsilon)$ is even in $\epsilon$ and monotonically increasing in $|\epsilon|$. 
\end{comment} 

\subsection{Observables}

In this section we compute the free energy of the dual three-dimensional SCFTs, the flavor central charge for various global symmetries and the dimensions of heavy operators dual to D2-branes localized in $AdS_4$.  

\subsubsection{Free Energy}

The free energy is given as
 \begin{equation}
 \mathcal{F} = \frac{\pi L^2_{AdS_4}}{2G_{4}} = \frac{16 \pi^3}{(2\pi \ell_s)^8}  \int e^{8A-2\phi} \Vol(M_6).
 \end{equation} In the second equality, we have used the choice $L_{AdS_4}=1$. 
 A short computation yields 
 \begin{equation}
 \mathcal{F}_{g\neq 1} =\frac{8\pi }{5 }\frac{ 2(1-g) N^{5/2}} {\kappa n_0^{1/2}} F_z(z,\kappa), \qquad F_z(z,\kappa)=\frac{(|z^2-\kappa^2|)^{3/2}(\sqrt{\kappa^2 +8z^2} -\kappa)}{(|4z^2 -\kappa^2 + \kappa \sqrt{\kappa^2 + 8z^2}| )^{3/2}}.   \label{freeener}
 \end{equation}  The free energy when the Riemann surface is a torus ($g=1$) can be obtained from a delicate limit where
 \begin{equation}
  \kappa \to 0, \qquad \frac{1-g}{\kappa} \to \frac{1}{\alpha} \to 4\pi, \qquad z = (q-p) \alpha = \frac{\hat{z}}{4\pi}, \qquad \hat{z} \in \mathbb{Z}.
  \end{equation}
 In this limit, we obtain the free energy for compactifications on the torus as
 \begin{equation}
 \mathcal{F}_{g=1} = \frac{4 \sqrt{2}\pi }{5 }\frac{ N^{5/2}} {n_0^{1/2}} |\hat{z}|.  
 \end{equation}
 
 We observe the expected scaling of the free energy in terms of $N$ and $n_0$ due to the five-dimensional origin of the dual SCFTs.  The dependence on the twist parameter is similar to central charge results for four-dimensional SCFTs when a-maximization \cite{Intriligator:2003jj} is needed to fix the superconformal R-symmetry \cite{Bah:2012dg}.  This dependence reflects the mixing of $U_\mathcal{M}(1)$ with $U_\mathcal{R}(1)$ to yield the superconformal R-symmetry for the three-dimensional theory.  This is an indication that $\mathcal{F}$-extremization \cite{Jafferis:2010un} will be necessary to fix the R-symmetry and scaling dimensions of protected operators in the dual SCFTs.

\begin{figure}
\centering
\includegraphics[scale=.5]{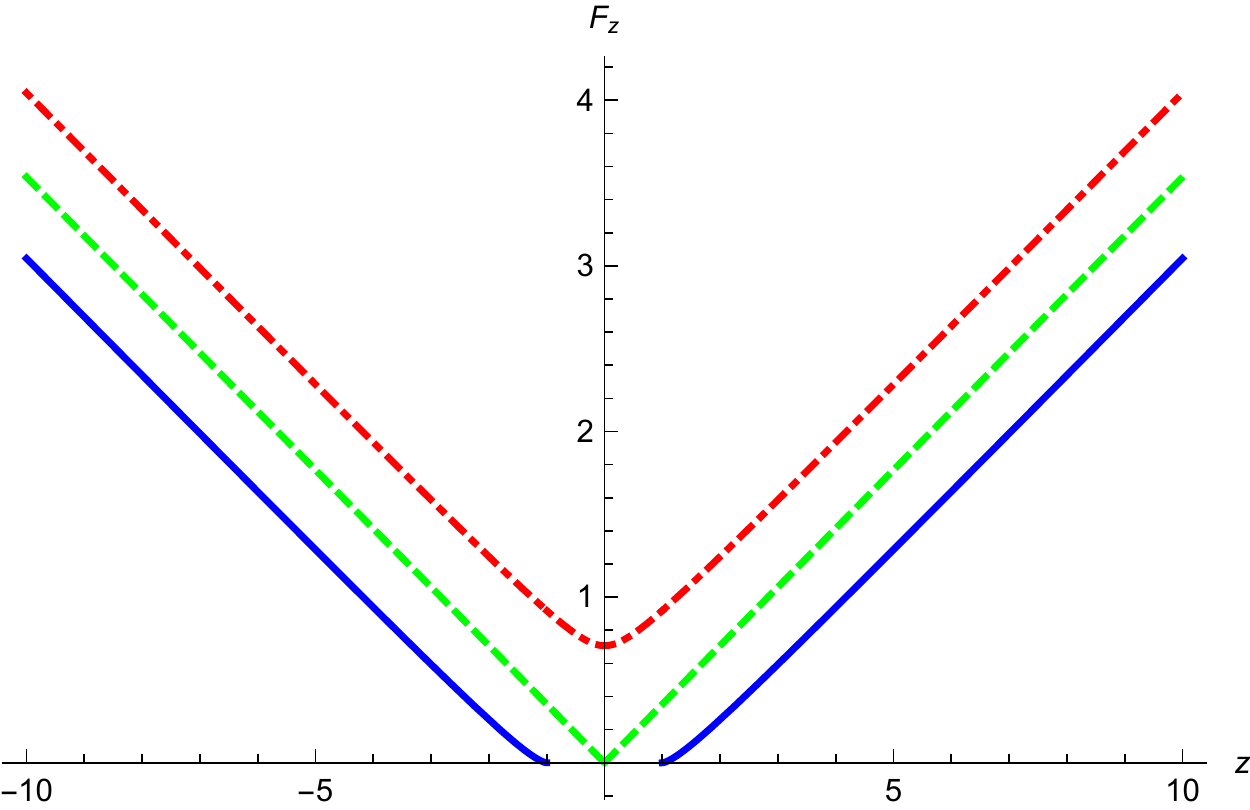}
\caption{The $z$ dependence of the free energy, for fixed $g$, $N$ and $n_0$, is plotted for $\kappa=-1$ (red and dot-dashed curve), for $\kappa=0$ (green and dashed curve), and for $\kappa=1$ (blue and solid curve).  Even though $F_z$ is plotted as a continuous function of $z$, only quantized values of $z$ correspond to SCFTs, as given in \eqref{pqpar} for fixed genus $g$.  }\label{Fig:Fz}
\end{figure} 

In Figure \ref{Fig:Fz} we plot the twist dependence of the free energy.  When $\kappa=-1$, we expect a SCFT for every integer choices of $(p,q)$ since the free energy is finite for all choices of $z$.  In particular there is a SCFT when $z=0$, i.e. when the $U_\mathcal{M}(1)$ is not twisted; the flavor symmetry is enhanced to $SU_\mathcal{M}(2)$.  The theory at $(\kappa=-1,z=0)$ was also studied in holography by using the six-dimensional $F(4)$ gauged supergravity \cite{Romans:1985tw}, in \cite{Nunez:2001pt,Karndumri:2012vh,Bobev:2017uzs}.  These authors also constructed the holographic RG flow to the $z=0$ theory from the six-dimensional Brandhuber-Oz solution \cite{Brandhuber:1999np}.  In \cite{Bobev:2017uzs} the authors also computed the free energy and their answer indeed matches with \eqref{freeener} in the $z=0$ and $\kappa=-1$ limit.  

The case with $\kappa=-1$ contrasts with the torus and sphere cases where the free energy from the supergravity vanishes in the former and is negative in the latter for $z=0$, as depicted in Figure \ref{Fig:Fz}.  The proper interpretation of this result in the case of the torus is that if there is a SCFT when $z=0$, the coefficient of the $N^{5/2}$ scaling vanishes. However the coefficient of $N^2$ and perhaps $N^{3/2}$ may be non-zero.  Such a theory would preserve eight supercharges.  One may be able to study its holographic dual in a type IIA background with O-planes after T-duality on the torus.  It should be studied as a D2-brane theory.  

The leading $N^{5/2}$ coefficient is negative in the case of the sphere with $z=0$.  This is a strong indication that there is no SCFT when the five-dimensional theory is compactified on the sphere while preserving the $SU_\mathcal{M}(2)$ symmetry.  It is interesting to wonder whether such a three-dimensional fixed point can exist when part of the $E_{N_f+1}$ symmetry is twisted.  When $|z|=1$ the coefficient of the leading $N^{5/2}$ term vanishes.  One may still expect a SCFT with an $N^2$ or $N^{3/2}$ scaling, but further investigation is also needed here.  Three-dimensional SCFTs are expected for $|z| \geq 2$.

\subsubsection{Flavor Central Charge}

The three-dimensional theories admit an $E_{N_f+1}$ flavor symmetry.  The flavor central charge for the $SO(2N_f) \subset  E_{N_f+1}$, which is realized by the gauge fields living on the nine-dimensional world-volume of the $N_f$ D8-branes located at $\mu_0 = 0$, can be computed easily in the supergravity solution \cite{Freedman:1998tz} (See \cite{Chang:2017mxc} for a nice discussion in the case of five-dimensional SCFTs).

The D8-brane action that couples to massive type IIA supergravity reads
\begin{equation}
S_{{\rm D}8} = - \frac{\mu_8}{\sqrt{2}\kappa_{10}} \int d^9x \sqrt{-g_9}
\left[ e^{-\phi} (\pi \ell^2_s)^2 {\rm Tr}(g^{\mu\rho}_9 g^{\nu\sigma}_9 \mathcal{F}_{\mu\nu} \mathcal{F}_{\rho\sigma}) + \dots \right]
\end{equation}
where $2\kappa_{10}^2 = (2\pi)^7 \ell_s^8$ and $\mu_8^2 = (2\pi)^{-9} \ell_s^{-10}$ are the gravitational coupling and D8-brane charge, respectively. The metric $g_9$ is the induced metric on the wordvolume of the D8-branes from the ten-dimensional metric near the O8/D8 locus.

Dimensionally reducing the D8-brane action on the internal space yields the following effective four-dimensional action
\begin{equation}
S_4 = -\frac{1}{2e^2}  \int d^4x \sqrt{-g_4} {\rm Tr}(\mathcal{F}_{\mu\nu} \mathcal{F}^{\mu\nu}) + \dots 
\end{equation} where $e$ is the gauge coupling of the gauge theory in $AdS_4$.  After the reduction, we obtain
\begin{equation}
\frac{1}{e^2} =  \frac{\pi}{2(2\pi\ell_s)^4} \frac{1}{n_0} \frac{e^{2\nu} \Vol(\Sigma_g)}{1-\epsilon^2}.  
\end{equation}  The flavor central charge \cite{Freedman:1998tz,Chang:2017mxc} is given as
\begin{align}
C_J^{SO(2N_f)} = \frac{2^6}{e^2} %&= \frac{64 \sqrt{2}}{3\pi} \frac{1-g}{\kappa} \frac{N^{3/2}}{n_0^{1/2}} (1-\epsilon^2)^{1/2} e^{2\nu} \\
&= \frac{32 }{3\pi} \frac{1-g}{\kappa} \frac{N^{3/2}}{n_0^{1/2}} \frac{(|z^2-\kappa^2|)^{1/2} (\sqrt{\kappa^2 + 8z^2} - \kappa)}{(|4z^2 -\kappa^2 + \kappa \sqrt{\kappa^2 + 8z^2}|)^{1/2}}.
\end{align}

Naively the flavor central charge for $U_I(1) \subset E_{N_f+1}$ should come from the effective $U(1)$ gauge symmetry from the $C_1$ potential associated to $F_2$ RR-flux.  However in the massive IIA background, the equation of motion of the $C_1$ potential involves both the NS-NS $B$ field and the $F_4$ RR-flux.  It is inconsistent to turn on an $AdS_4$ gauge field from $C_1$ without turning on a contribution from the NS-NS $B$ field and from the $C_3$ potential.  Such truncation is more involved and will not be considered here.  This point is also discussed nicely in \cite{Chang:2017mxc}.  

This issue about the $U_I(1)$ is also related to why we cannot readily turn on a twist parameter for the symmetry.  Such a twist parameter is realized in supergravity by an $F_2$ flux along the Riemann surface.  However in the analysis of $AdS_4$ solutions of \cite{Passias:2018zlm}, it is clear that one cannot turn on $F_2$ without also turning on the RR six-form flux, $F_6$, and the NS-NS three-form flux $H_3$.  It would indeed be interesting to explore such twist, and we leave it to future work.

\subsubsection{D2-brane Operators}  

We can study a class of $\frac{1}{2}$-BPS operators of the three-dimensional SCFTs, dual to D2-branes localized in $AdS_4$.  Such operators can be understood in gravity by considering a probe D2-brane as a point particle $AdS_4$ wrapping a two-cycle in $M_6$.  In particular we will need to solve the $\kappa$-symmetry conditions for D2-branes in the background given by \eqref{pqsol}.  The analysis for these D2-branes is analogous to the analysis of M2-brane probes in B$^3$W solutions \cite{Bah:2012dg,Bah:2013wda} (please refer to these papers for more details).  We review the $\kappa$-symmetry conditions in appendix \ref{kappasym}.  

The primary effect of $\kappa$-symmetry is to determine the possible locations for a probe brane that preserves some of the supersymmetry of the background space-time.  For the case of interest, the world-volume time direction is trivially identified with the $AdS_4$ time direction.  The two-cycle in $M_6$ is given by a calibation condition that fixes the volume of the probe as 
\begin{equation}
\Vol({\rm D}2)_{M_6} = \left. \Phi \right|_{{\rm D}_2},
\end{equation} where the left-hand side is the volume form of the two-cycle wrapped by the D2-brane probe, and the right-hand side corresponds to the pullback of a globally defined two-form of the geometry onto the world-volume of the probe.  The calibration condition provides a set of differential equations whose solutions yield a two-dimensional curve in $M_6$.  In the case of the D2-brane probe (see appendix \ref{kappasym} for details), the global form is given as
\begin{align}
\Phi &=ye^{-2A}\left[ \frac{dy}{2y} \wedge \eta_\psi + \frac{e^{-4A}}{y F_0^2} J \right]\\
&=\left(3t H\right)^{1/2} \left[ H J  -\frac{dt}{6t} \wedge \left(\eta_t - A_u \eta_u\right) \right]
\end{align} where $J$ and $(t,u)$ are given in \eqref{Jdef} and \eqref{tucoord}. Furthermore, $\kappa$-symmetry dictates the D2-brane to be located at the center of $AdS_4$.

Given the calibration condition, the scaling dimension of the dual operator, $\mathcal{O}_{{\rm D}2}$, in the SCFT is given by the effective mass\footnote{More precisely, the scaling dimension is the effective mass in units of the $AdS_4$ radius, $L_{AdS_4}$.  This is fixed to $1$ in our conventions.} of the point particle in the bulk of $AdS_4$.  This is obtained by reducing the DBI action for the D2-brane in the geometry:\footnote{The mass of the point particle from the D2-brane is extracted from the action \begin{equation} S= -\mu_{D2} \int \widetilde{\Vol}(D_2) e^{-\phi} \nonumber
\end{equation} where $\mu_{D_2} =(2\pi)^{-2} (\ell_s)^{-3}$ is the tension of the D2-brane and $\widetilde{\Vol}({\rm D}_2)$ is the volume form of the brane obtained from the pullback of the full ten-dimensional metric.}
\begin{equation}
\Delta(\mathcal{O}_{{\rm D}2} ) = \frac{2\pi}{(2\pi \ell_s)^{3} } \int \Vol({\rm D}2)_{M_6} e^{-\phi} \left( y F_0^2 H\right)^{-3/4}.
\end{equation}

The probe must preserve the R-symmetry circle which is dual to the one-form $(\eta_t -A_u \eta_u)$.  The two-cycle calibrated by $\Phi$ can either sit at places where the R-symmetry circle shrinks or it must be extended along it.  To study these configurations more carefully, we write the $M_6$ metric as
\begin{align}
ds^2(M_6) &=e^{2\nu} ds^2(\Sigma_g) + \frac{1}{3} \frac{1}{q(\theta) }\frac{d\mu_0^2}{1-\mu_0^2} +\frac{2H(1-\mu_0^2)}{1-\epsilon^2} ds^2(M_3)\\
ds^2(M_3) &= q(\theta)\left(d\theta - \epsilon \frac{\sin(2\theta)}{2q(\theta)} \frac{\mu_0 d\mu_0}{1-\mu_0^2 }\right)^2 + \frac{\sin^2(\theta) \cos^2(\theta)}{q(\theta)} \eta_u^2 + \frac{(1-\epsilon^2) q(\theta)}{4} \left(\eta_t - A_u \eta_u\right)^2 \nonumber
\end{align} where $A_u = -\cos(2\theta)/q(\theta)$.

\subsubsection*{D2-brane on $\Sigma_g$}
It is clear from the metric of $M_6$ that the R-symmetry circle shrinks only at $\mu_0 =1$ where the full three-manifold $M_3$ also shrinks.  In this case, the brane can only wrap the Riemann surface.  The scaling dimension for the dual operator $\mathcal{O}_{{\rm D}2^\Sigma}$ is
\begin{align}
\Delta(\mathcal{O}_{{\rm D}2^\Sigma}) %&= \frac{2\pi}{(2\pi \ell_s)^{3} } \frac{\sqrt{y}}{\sqrt{H}}  \frac{4\pi(1-g)}{\kappa} e^{2\nu} \\
%&=\frac{2(2\pi)^{2}}{(2\pi \ell_s)^{3} }\left(\frac{3}{2}\right)^{2/3} F_0^{-1/3}  \frac{(1-g)}{\kappa} e^{2\nu} \\
%&= 2(2\pi)^{2} \left(\frac{3}{2}\right)^{2/3} n_0^{-1/3} (2\pi \ell_s)^{-8/3}  \frac{(1-g)}{\kappa} e^{2\nu} \\
&= \frac{2(1-g)}{\kappa} N \frac{(z^2-\kappa^2) \left(\sqrt{\kappa^2 + 8z^2} -\kappa\right)}{4z^2 -\kappa^2 + \kappa \sqrt{\kappa^2 + 8z^2}}.
\end{align} When $\kappa =0$ we can take the limit $\frac{1-g}{\kappa} \to 4\pi$ and $z= \frac{\hat{z}}{4\pi}$. 

Note that this operator does not scale with $n_0$ and is not sensitive to the number of flavors, $N_f$.  This is a strong indication that the operator is neutral under the $E_{N_f+1}$ flavor symmetry.  It is also interesting to note that the scaling dimension vanishes for $(z=0,\kappa=0)$ and for $(z=1, \kappa =1)$.  As we saw before, the leading $N^{5/2}$ coefficient of the free energy also vanishes.  Unlike the free energy, however, one may expect that if there exist SCFTs for these twists, the scaling dimension of the operator dual to the D2-brane probe will be independent of $N$.  This is consistent with the fact that in these limits, the size of the Riemann surface is going to zero and thus the D2-brane is wrapping a collapsing cycle.

\subsubsection*{D2-brane on $\eta_t-A_u\eta_u$}

The probe can also wrap the R-symmetry circle with dual one-form $(\eta_t -A_u \eta_u)$.  In this case, the calibration condition reduces to
\begin{equation}
\Vol({\rm D}2)|_{M_6} = - \left. \left(3t H\right)^{1/2} \frac{d\mu_0}{3\mu_0} \wedge \left(\eta_t - A_u \eta_u\right) \right|_{{\rm D}2}, \qquad \left. J\right|_{{\rm D}2} =0.  
\end{equation}  The calibration condition implies that the D2-brane probe must also be extended along the $\mu_0$ interval.  At $\mu_0 =1$, the R-symmetry circle shrinks and the D2-brane smoothly caps off.  However at $\mu_0=0$ the D2-brane must end on a D8-brane in order for the configuration to be consistent.  The operator dual to such a configuration must be charged under the $E_{N_f+1}$ flavor symmetry.  These are a class of operators that exist for the three-dimensional SCFTs which have no analogue in the field theory dual of the B$^3$W solutions \cite{Bah:2012dg}.  However when there is a puncture on the Riemann surface,  there are additional M5-brane sources; thus there is a M2-brane probe that is anologous to the D2-brane probe where the M2-brane ends on the M5-brane \cite{Bah:2013wda,Gaiotto:2009gz}.  In those cases, the operators dual to the M2-brane probe correspond to baryonic operators charged under the flavor group induced by the M5-brane sources.  From these considerations, one can expect the operators dual to the D2-brane probe to be charged under the $E_{N_f+1}$ flavor symmetry.    

We can consider solutions where the D2-brane sits at a point on the Riemann surface and has a profile along the $(\mu_0, \theta)$ directions.  It is convenient to consider the world-volume coordinates as $(\sigma, \varphi)$.  The embedding of the brane is given by the functions 
\begin{equation}
\left(\theta(\sigma), \mu_0(\sigma)\right), \qquad \left(\phi_+(\varphi), \phi_-(\varphi)\right), \qquad x_i = \mbox{constant}.
\end{equation}  The embedding functions are determined by the calibration conditions, which reduce to
\begin{equation}
\sin(2\theta) \left((1-\epsilon)\frac{d\phi_+}{d\varphi} - (1+\epsilon) \frac{d\phi_-}{d\varphi} \right)=0, \qquad  \frac{d\theta}{d\sigma} - \epsilon \frac{\sin(2\theta)}{2q(\theta)} \frac{\mu_0 }{1-\mu_0^2} \frac{d\mu_0}{d\sigma}  =0.
\end{equation}  

There are three types of solutions to consider.  There is a solution, ${\rm D}2^+$, where the probe sits at $\theta =0$ and thus must wrap the $\phi_+$ circle since the $\phi_-$ circle shrinks.  There is also a similar solution, D2$^-$ where the probe sits at $\theta=\frac{\pi}{2}$ and wraps the $\phi_-$ circle since the $\phi_+$ circle shrinks.  In both cases, the probe is simply extended along $\mu_0$, i.e. we can take $\mu_0 = \sigma$.  The scaling dimensions for the probes are 
\begin{align}
\Delta({\rm D}2^\pm) %&= \frac{1}{3} \left(\frac{3}{2} \right)^{5/3} (2\pi)^2 \frac{n_0^{-1/3}}{2a_\pm} (2\pi \ell_s)^{-8/3} \\
%&=2 a_\mp N \\
&= N \left(1 \mp \frac{\kappa + \sqrt{\kappa^2 + 8z^2}}{4z}\right).
\end{align}

The last solution, D2$^0$, is less trivial.  For this solution we have $\eta_u=0$ and the D2-brane has a profile along $\theta$ as well as $\mu_0$. The circle coordinates are
\begin{equation}
\phi_\pm = c_\pm - m_\pm \varphi, \qquad \mbox{with} \qquad \frac{m_+}{m_-} = \frac{1+\epsilon}{1-\epsilon}.
\end{equation} The length of the D2-brane must be finite on the torus spanned by $(\phi_+, \phi_-)$.  This implies that the winding parameters $m_\pm$ must be rational. We can pick them to be co-prime integers without loss of generality.  These configurations can exist, then, only when $\frac{1+\epsilon}{1-\epsilon}$ is rational or, equivalently, when the free energy is rational.\footnote{The family of solutions when $\epsilon$ is rational are interesting to study.  In particular one can show that the three-manifold $M_3$ at fixed $\mu_0$ is a Seifert manifold where the $(\theta, \phi_u)$ directions make a sphere with orbifold fixed points at the north and south poles.  The $\eta_t$ circle is fibered over the sphere in such way to smooth out the orbifold singularities in $M_3$ (see \cite{Bah:2013wda} for more details).}  

For the D2$^0$ solution, there is a nontrivial profile along the $(\mu_0, \theta)$ directions given by the curve
\begin{equation}
\mu_0^2 = 1 - C (\cos(\theta))^{\frac{1-\epsilon}{\epsilon}} (\sin (\theta))^{\frac{-1-\epsilon}{\epsilon}}.
\end{equation}  The integration constant $C$ is fixed by the angle $\theta_0$ where the D2-probe intersects the D8-brane.  The scaling dimension for this solution is 
\begin{equation}
\Delta ({\rm D}2^0) = N(m_+ + m_-) =  m_\pm \Delta({\rm D2}^\pm).  
\end{equation}

When the D2$^0$ probe sits at $\theta=0$ it behaves as $m_+$ for the D2$^+$ probes, and when it sits at $\theta= \frac{\pi}{2}$, it behaves as $m_-$ for the D2$^-$ probes.  This fractionalization is reflected in the scaling dimensions of the dual operators.

%%%%%%%%%%%%%%%%%%%%%%%%%%%%%%%%%%%%%%%%%%%
%%%%%%%%%%%%%%%%%%%%%%%%%%%%%%%%%%%%%%%%%%%

\section{Discussion and Conclusion}

In this paper we have discussed the topological twist of five-dimensional Seiberg SCFTs \cite{Seiberg:1996bd} with $SU_\mathcal{M}(2)\times E_{N_f+1}$ flavor symmetry on a punctured Riemann surface.  The primary question we hoped to address was whether any such twist can lead to three-dimensional SCFTs that we can label and describe by the Riemann surface, similar to the class $\mathcal{S}$ program.   We explicitly constructed the holographic duals of such SCFTs when the Riemann surface is smooth and has no punctures, and when the $U(1)$ subgroup of $SU_\mathcal{M}(2)$ is twisted along with the R-symmetry.  The geometric origin of both of these symmetries allows for a more straightforward study of their twists in holography.  The setup and construction of these geometries is exactly analogous to the twisted compactifications of the six-dimensional $(2,0)$ SCFTs on a Riemann surface, as described in \cite{Gaiotto:2009hg,Gaiotto:2009gz,Bah:2012dg}.

The general supergravity system we derived is a candidate for encompassing the gravitational duals of all possible three-dimensional SCFTs that can appear in the compactifications of the five-dimensional Seiberg theories on punctured Riemann surfaces, when the R-symmetry and the $U(1)$ subgroup of $SU_\mathcal{M}(2)$ are twisted.  The computation of the free energy and its recent matching with \cite{Crichigno:2018adf, Hosseini:2018uzp} provides strong evidence for such a duality.  It will be important to construct the holographic RG flows from the $AdS_6$ duals of Seiberg theories to the solutions in this paper.  This is a subject that is of great import to the authors and will be studied in the future.  

It is interesting to study how to describe the SCFTs with punctures within our system.  Indeed, this question has been studied for the holographic duals of twisted compactifications of six-dimensional $(2,0)$ SCFTs \cite{Gaiotto:2009gz,Bah:2013qya,Bah:2015fwa}.  The system derived is analogous to the $AdS_5 \times M_6$ solutions there.  The supergravity description of punctures will thus be the same. However, the nature of the sources that lead to the punctures will be different.  This is a problem we hope to study in the future.  

One way to understand the sources that correspond to punctures is to consider the set of BPS probes that we can turn on in the solutions described above.  Such an analysis was conducted for the $AdS_5$ solutions of B$^3$W in \cite{Bah:2013wda}.  In particular there exist D4-brane probes that are localized on the Riemann surface.  A collection of such defects can allow for a more extensive exploration of the set of possible punctures.  There can also be more general defect objects from other types of brane sources.  

An important aspect of the class $\mathcal{S}$ theories is that they are organized by a Topological Quantum Field Theory (TQFT) on the punctured Riemann surface \cite{Gaiotto:2009hg}.  Such structure was understood in the case of $\mathcal{N}=1$ class $\mathcal{S}$ theories at the level of their superconformal index in \cite{Beem:2012yn}.  We expect the three-dimensional SCFTs described above to also admit a TQFT structure.  This is a prediction from the holographic system above since it is exactly analogous to the holographic duals of $\mathcal{N}=1$ class $\mathcal{S}$.  This can be realized by computing the partition function of the five-dimensional Seiberg SCFTs on $S^3 \times \Sigma_{g,n}$.  The reduction of partition function on the Riemann surface should exhibit a TQFT.  

Further exploration of the SCFTs above will yield great insight into the space of $\mathcal{N}=2$ three-dimensional SCFTs.  In particular, it would be interesting to explicitly construct the dual SCFTs in terms of ingredients from three-dimensional QFTs.  Although one might expect a generic SCFT in this class to be strongly-coupled, one may be able to identify basic building blocks similar to $T_N$ theories for four-dimensional class $\mathcal{S}$ theories \cite{Gaiotto:2009we,Gaiotto:2009gz}.  

One can also consider the twisted compactifications of the very large class of five-dimensional SCFTs such as the ones discussed in \cite{Intriligator:1997pq} and from the more complete, recent, classification in \cite{Jefferson:2018irk}.  Holographic duals can constructed by also considering the twisted compactifications of five-dimensional SCFTs when the D4-branes are at orbifold fixed points such as in \cite{Bergman:2012kr}.  In these cases, one explores a larger class of twists due to the proliferation of the flavor symmetry.  Along the same lines, it would interesting to understand the SCFTs and their holographic duals when the $E_{N_f+1}$ flavor symmetry is twisted.  

Indeed this is the tip of the iceberg for a much more interesting and elaborate story about three-dimensional SCFTs and five-dimensional SCFTs similar to the story of four-dimensional SCFTs from six-dimensional $(2,0)$ and $(1,0)$ theories.    

{\bf Note added:} After this work appeared on arXiv, the work of \cite{Crichigno:2018adf, Hosseini:2018uzp} appeared, which reproduced the expression \eqref{FEintro} for the free energy, by calculating the partition function of the five-dimensional field theory on a Riemann surface.

\acknowledgments

The authors thank Alessandro Tomasiello and Emily Nardoni for useful discussion.   The work of AP is supported
by the Knut and
Alice Wallenberg Foundation under grant Dnr KAW 2015.0083.

\appendix
\section{Derivation of the System}\label{derivation}
\subsection*{General System}
The general form of the metric of a $\mathcal{N}=2$ supersymmetric $AdS_4$ solution in massive type IIA supergravity, with $SU(2)$-structure on the internal manifold is \cite{Passias:2018zlm} 
\begin{equation}
ds^2_{10} =e^{2A} \left[ ds^2(AdS_4) + \frac{dy^2}{e^{4A}-y^2} + \frac{e^{4A} -y^2}{4e^{4A}} \left(d\psi + \rho\right)^2 + \frac{y}{\ell^2 + y^2 F_0 e^{4A}} ds^2(N_4) \right],
\end{equation}
as in equation (\ref{ogmetric}). The equations dictated by supersymmetry, governing the structure of the four-dimensional subspace $N_4$, can be expressed in terms of the real two-form $J$ and holomorphic two-form $\Omega$ as
\begin{equation}\label{Asystem2}
\begin{alignedat}{8}
\partial_\psi J &{}= 0  , & \qquad &
\partial_yJ &{}={}& \frac{1}{2} (F_0^2 y^2 + \ell^2 y^{-2}) d_4 \rho \ , & \qquad &
d_4J &{}={}& 0  , \\
\partial_\psi \Omega &{}= 0  , & \qquad &
\partial_y \Omega &{}={}& - \frac{1}{2} (F_0^2 y^2 + \ell^2 y^{-2}) T \Omega  , & \qquad &
d_4 \Omega &{}={}& iP \wedge \Omega ,
\end{alignedat}
\end{equation}
where
\begin{align}
P &\equiv - \rho + i \frac{2 e^{4 A} (\ell^2 + F_0^2 y^4) }{(e^{4A}-y^2)(\ell^2 + F_0^2  e^{4A}  y^2)} d_4 A , \\
T &\equiv \frac{\partial_y(e^{4A}y^2)}{(e^{4A}-y^2)(\ell^2 + F_0^2  e^{4A}  y^2)}.
\end{align}
The purpose of this appendix is to reduce these equations for our problem of interest. We restrict to solutions with $\ell=0$, setting all but the four-form flux to zero. This is assumed for the remainder of this appendix.

$d_4 \Omega = iP \wedge \Omega$ implies that the almost complex structure defined by $\Omega$ is independent of $\psi$ and $y$, and is integrable on $N_4$.
Additionally, $d_4J = 0$ implies that the metric on $N_4$ is locally a family of K\"{a}hler metrics parametrized by $y$. 

A series of general conditions follow from the system given in \eqref{Asystem2} which will be useful in reducing the supersymmetry equations for the specific $U(1)^2$ ansatz of interest here. The condition for $\partial_y \Omega$ determines the dependence of the $N_4$ volume on $y$:
\begin{equation}
\partial_y \log \sqrt{g^{(4)}} = (F_0^2 y^2 + \ell^2 y^{-2}) T  .
\end{equation}
The independence of the complex structure from $y$ leads to
\begin{equation}
(\partial_yJ)^+ = \frac{1}{2} \partial_y \log \sqrt{g^{(4)}} \,J ,
\end{equation}
where the plus superscript denotes the self-dual part of a 2-form on $N_4$. Combining with the second equation of \eqref{Asystem2} we obtain
\begin{equation}\label{d4rhoplus}
(d_4 \rho)^+ = - TJ .
\end{equation}
The system \eqref{Asystem2} also yields:
\begin{equation}\label{Aint}
d_4 \rho \wedge \Omega = 0  , \qquad \left[\frac{8y}{(e^{2A}-e^{-2A}y^2)^2} d_4 A + i \partial_y \rho \right] \wedge \Omega = 0  .
\end{equation}

Turning now to the other fields, the dilaton is given by
\begin{equation}
e^{2\phi} = \frac{e^{2A}}{y^2 F_0^2} .
\end{equation}
Solutions are supported by a four-form flux which can be written
as
\begin{equation}
F_4 = -\left(\frac{1}{F_0}  J + \frac{1}{2} F_0 y^2 dy \wedge D\psi \right) \wedge \alpha_2 - \frac{1}{2} \frac{e^{-4A}}{F_0^3 y^2} J \wedge J ,
\end{equation}
where
\begin{equation}
\alpha_2 \equiv - \frac{1}{2} d (e^{-4A} y (d\psi + \rho)) + \frac{1}{2}  y^{-1} d\rho.
\end{equation}

\subsection*{Ansatz with $U(1)^2$ Isometry} 
We can write the ansatz of equation (3.7)
\begin{equation}
ds^2 (N_4) = e^{2W}\left(dx_1^2 + dx_2^2\right) + e^{2Z} \left[\left(d\tau+V^R \right)^2 + e^{2C} \left(d\phi+ V^I\right)^2 \right]
\end{equation}
also as
\begin{equation}
ds^2_4 (N_4) = e^{2W} \epsilon_1 \bar{\epsilon}_1 + e^{2Z} \epsilon_2 \bar{\epsilon}_2,
\end{equation} where
\begin{align}
\epsilon_1 &= dx_1 + i dx_2, \qquad \epsilon_2 = \eta_\tau + i e^C \eta_\phi \\
\eta_\tau  &= d\tau + e^C V^R, \qquad \eta_\phi = d\phi + V^I,
\end{align} and the one forms $V^R$ and $V^I$ have legs along the $x$-directions of the Riemann surface.  It is useful to split the exterior derivative as
\begin{equation}
d_4 = d_2 + \eta_\tau \partial_\tau + d\phi \partial_\phi, \qquad d_2 = \hat{d} -e^{C} V^R \partial_\tau,
\end{equation} where $\hat{d}$ is the exterior derivative along the $x$-directions.

With this ansatz we can write our compatible K\"{a}hler and almost complex structure as
\begin{align}
J &=e^{2W} d R_2  +  e^{2Z+C} \eta_\tau \wedge \eta_\phi, \qquad dR_2 = dx_1 \wedge dx_2 \\
\Omega &= e^{W+Z} e^{i q \psi + i p \phi}  \epsilon_1 \wedge \epsilon_2 =  e^{W+Z} e^{i q \psi + i p \phi} \left(\Omega_0^R + i \Omega_0^I \right) \\
\Omega_0^R &= dx_1 \wedge \eta_\tau - e^C dx_2 \wedge \eta_\phi \\
\Omega_0^I &= e^C dx_1 \wedge \eta_\phi + dx_2 \wedge \eta_\tau.  
\end{align} where $p$ and $q$ are the charges of $\Omega$ under the $\phi$ and $\psi$ $U(1)$'s. If we define
\begin{equation}
e^{-4A} \equiv \frac{1}{y^2} \left( 1 - \cos^2(\zeta) \right), \qquad \eta_{\psi} \equiv d\psi + \rho.  
\end{equation} 
the full metric can be rewritten as
\begin{equation}
ds^2_{10} = e^{2A} \left[ ds^2(AdS_4) + \frac{1}{4} \cos^2(\zeta) \eta_\psi^2 + e^{-4A} \frac{dy^2}{\cos^2(\zeta)} + \frac{1}{y F_0^2} e^{-4A} ds^2(N_4) \right].
\end{equation} 
Before proceeding to the reduction of the supersymmetry equations for this ansatz, note that the Hodge star conventions will be
\begin{align}
\star_2 dx_1&= - dx_2, \qquad \star_2 dx_2=  dx_1, \qquad \star_2 \epsilon_1 = i \epsilon_1, \qquad \star_2 \bar{\epsilon}_1 = i \bar{\epsilon}_1 \\
\star_\tau \eta_\tau &= - e^C \eta_\phi, \qquad \star_\tau \eta_\phi e^C = \eta_\tau.
\end{align}

\subsection*{Reducing the $\partial_y \Omega$ Equation}
With $\ell=0$, this equation is
\begin{equation}
\frac{\partial_y \Omega}{\Omega} = - \frac{\partial_y(e^{4A}y^2)}{2 e^{4A}(e^{4A}-y^2)},
\end{equation}
which implies that 
\begin{equation}
y \partial_y \log e^{W+Z}+y \partial_y  \log \epsilon_1 \wedge \epsilon_2 =-4 \tan^2 \zeta- y \partial_y  \log\left( \cos^2 \zeta \right).
\end{equation}
This leads to two conditions. First, that
\begin{equation}
y \partial_y  \log \epsilon_1 \wedge \epsilon_2=0,
\end{equation}
and second, that
\begin{equation}\label{zetaeq}
y\partial_y \log \left[e^{2W+2Z} \cos^2(\zeta) \right] =-4\tan^2(\zeta).  
\end{equation}
Note that the complex two-form equation $\partial_y(\epsilon_1 \wedge \epsilon_2)=0$ yields two conditions. Together they imply
\begin{equation}
\partial_y C =0, \qquad \epsilon_1 \wedge \partial_{y} \left(V^R+ i V^I \right) =0.
\end{equation} 
Thus we can fix $C=0$ in the ansatz without loss of generality.  Furthermore, we note that $\partial_y \left(V^R + i V^I \right) $ is holomorphic.  We can then write
\begin{equation}
V^{I} = V_0 - \star_2 V^R, \qquad \partial_y V_0 =0.  
\end{equation} 
Finally, note that \eqref{zetaeq} admits the following solutions:
\begin{equation} \label{warpsol}
\cos^2(\zeta) = - \frac{4}{y\partial_y \Lambda}, \qquad e^{2W + 2Z} = - \frac{1}{4} y^5 \partial_y e^\Lambda, 
\end{equation}
where $\Lambda$ is a scalar function.
\subsection*{Reducing the $d_4 \Omega$ Equation}

The next equation to solve is 
\begin{equation}
d_4 \Omega = i P \wedge \Omega.  
\end{equation} 
First note that
\begin{equation}
P = - \rho + i d_4 \log \left(\cos \zeta \right)=-\rho-\frac{i}{2}\frac{d_4(y \partial_y \Lambda)}{y \partial_y \Lambda}
\end{equation}
and thus
\begin{equation}
d_4 \Omega = \left[ d_4 (W+Z) + ip d\phi \right] \wedge \Omega - e^{ip \phi} e^{W+Z} \epsilon_1 \wedge d_4 \left(V^R + iV^I \right).  
\end{equation}  This relation yields two equations
\begin{equation}
\left[ \frac{1}{2} d_4 \Lambda + i (\rho+p d\phi) \right] \wedge \Omega =0, \qquad \epsilon_1 \wedge \partial_\tau \left(V^R + iV^I \right) =0. 
\end{equation} The second equation is solved by
\begin{equation}
V^{I} = V_0 - \star_2 V^R, \qquad \partial_\tau V_0 =0, 
\end{equation} and the first by
\begin{equation}\label{rhoeq}
\rho = - p d\phi -\frac{1}{2} \star_2 d_2 \Lambda + \frac{1}{2} \partial_\tau \Lambda \eta_\phi.  
\end{equation}

\subsection*{Reducing the $d_4 J$ Equation} 

When expanded, $d_4 J=0$ implies that
\begin{equation}
\eta_\tau \wedge (\partial_\tau e^{2W} dR_2-e^{2z} d_2 V^I)+(d_2 e^{2Z}-e^{2Z}\partial_\tau V^R)\wedge \eta_\tau \wedge \eta_\phi +e^{2Z} d_2 V^R \wedge \eta_\phi =0.
\end{equation}
Thus we have three equations:
\begin{align}
d_2 V^R &= 0 \label{dj1}\\
d_2 e^{2Z} &= e^{2Z} \partial_\tau V^R \label{dj2} \\
e^{2Z} d_2 V^I &= \partial_\tau e^{2W} dR_2.  \label{dj3}
\end{align} 

\subsection*{Reducing the $\partial_y J$ Equation}
The equation for $\partial_y J$ is
\begin{equation}
\partial_yJ =\frac{1}{2} F_0^2 y^2 d_4 \rho.
\end{equation}
Now we can evaluate this last equation using the solutions found above. Expanding $\rho$, using equation \eqref{rhoeq} on the right-hand side and $J$ on the left-hand side, we obtain:
\begin{align}
\partial_y e^{2Z} &= \frac{F_0^2 y^2}{4} \partial_\tau^2 \Lambda \label{yj1} \\
e^{2Z} \partial_y V^R &= \frac{F_0^2 y^2}{4} d_2 \partial_\tau \Lambda \label{yj2} \\
\partial_y e^{2W} dR_2 &=  \frac{F_0^2 y^2}{4}\left[\partial_\tau \Lambda d_2 V^I - d_2 \star_2 d_2 \Lambda \right] \label{yj3} \\
e^{2Z} \partial_y V^I &= \frac{F_0^2 y^2}{4} \left[\partial_\tau \Lambda \partial_\tau V^I - \partial_\tau \left(\star_2 d_2 \Lambda \right) \right]. \label{yj4}
\end{align} Note that the $\partial_y V^I$ equation is implied by the relation of $V^I$ with $V^R$ and the $\partial_y V^R$ equation.  

\subsection{Further Reduction} 

We are left with \eqref{dj1}-\eqref{dj3} and \eqref{yj1}-\eqref{yj4} to solve.
Note that equation \eqref{dj1} implies
\begin{equation}
V^R = d_2 \Gamma 
\end{equation} where $\Gamma$ is a scalar function.  So let us choose new coordinates $t,u$ to replace $y, \tau$, such that
\begin{equation}
y^3 F_0^2 = 3 t, \qquad \tau = -\Gamma \qquad \Longrightarrow \qquad  \eta_\tau = - \partial_u \Gamma du - \partial_t \Gamma dt
\end{equation}
and thus
\begin{equation}
\partial_\tau = - \frac{1}{\partial_u \Gamma} \partial_u, \qquad \frac{1}{y^2 F_0^2} \partial_y = \partial_t - A_u \partial_u,
\end{equation} where $A_u = \partial_t \Gamma/\partial_u \Gamma$. We can then write the solution to equation \eqref{dj2} as
\begin{equation}
e^{2Z} = - \frac{G}{\partial_u \Gamma}, \qquad d_2 G =0, 
\end{equation} for some function $G$. 
To proceed, define
\begin{equation} \label{dets}
g \equiv \partial_t \Lambda \partial_u \Gamma - \partial_t \Gamma \partial_u \Lambda, \qquad h \equiv \partial_t \widetilde{\Lambda} \partial_u \Gamma - \partial_t \Gamma \partial_u \widetilde{\Lambda}, \qquad \widetilde{\Lambda} \equiv \Lambda + \frac{4}{3} \ln t - \ln c_0
\end{equation} and $H \equiv h/(3tg)$, where $c_0$ is a parameter which we will later fix for convenience.  As will be evident later, $h$ and $g$ are determinants in the internal manifold, and $H$ an overall warp factor of the full metric. Now using the solution for $e^{2W+2Z}$ in \eqref{warpsol} and $e^{2Z}$ found above, we have
\begin{equation}
e^{2W} = \frac{3^{7/3}F_0^{-8/3} c_0}{4G} t g. e^{\widetilde{\Lambda}}  
\end{equation} 
Also note
\begin{equation}
e^{-4A} = y F_0^2 H.
\end{equation}
Next, since $V_0$ is independent of both $\tau$ and $y$, we can set it to zero without loss of generality, so $V^I=-*_2 V^R$. This reduces equation \eqref{dj3} as:
\begin{equation}
\partial_u e^{2W} dR_2 = -G d_2 \star_2 d_2 \Gamma, \qquad \Longrightarrow \qquad \partial_u e^{2W} = G \left(\partial_{x_1}^2 + \partial_{x_2}^2 \right) \Gamma.
\end{equation} 
Defining a new function $\tilde{G} \equiv \frac{4G \partial_t \Gamma -\partial_u \Lambda}{4 \partial_u \Gamma}$,  equations \eqref{yj1} and \eqref{yj2} become
\begin{equation}
\partial_u \tilde{G} = \partial_t G, \qquad d_2 \tilde{G} = 0.  
\end{equation} Finally, the last equation \eqref{yj3} becomes
\begin{equation}
\left(G\partial_t - \tilde{G} \partial_u \right) e^{2W} dR_2 = -\frac{G}{4} d_2 \star_2 d_2 \Lambda, \quad \Longrightarrow \quad \left(G\partial_t - \tilde{G} \partial_u \right) e^{2W} = \frac{G}{4} \left(\partial_{x_1}^2 + \partial_{x_2}^2 \right) \Lambda.
\end{equation} Before we write the metric, we write the one-forms for the $U(1)$'s as
\begin{equation}
-\frac{1}{2} \eta_\phi = \eta_u = d\phi_u + \frac{1}{2} \star_2 d_2 \Gamma, \qquad \eta_t = d\phi_t + \frac{1}{2} \star_2 d_2 \Lambda, \qquad -\eta_\psi = \eta_t - \frac{\partial_u \Lambda}{\partial_u \Gamma} \eta_u.  
\end{equation} The metric becomes
\begin{align}
ds^2_{10} &= \frac{H^{-1/2}}{\sqrt{y} F_0 } \left[ ds^2(AdS_4) +  H e^{2W} \left(dx_1^2 + dx_2^2 \right) + \frac{1}{4} H ds^2_4 \right] \\
ds^2_4 &= -\frac{g}{\partial_u \Gamma} dt^2 - 4G \partial_u \Gamma \left(du + \frac{\partial_t \Gamma}{\partial_u \Gamma} dt \right)^2 - \frac{16G}{\partial_u \Gamma} \eta_u^2 - \frac{4\partial_u \Gamma}{h} \left(\eta_t - \frac{\partial_u \Lambda}{\partial_u \Gamma} \eta_u \right)^2.
\end{align} From the equations of $G$ and $\tilde{G}$, we can remove them by making the coordinate transformation
\begin{equation}
dt' = dt, \quad du' = G du + \tilde{G} dt, \qquad G\partial_{u'} = \partial_u, \quad G \partial_{t'} = G \partial_t - \tilde{G} \partial_u.  
\end{equation} A convenient choice is then
\begin{equation}
G = \frac{1}{4}, \qquad \tilde{G} =0, \qquad \Longrightarrow \qquad \partial_u \Lambda = \partial_t \Gamma.  
\end{equation} We then have
\begin{equation}
\Lambda = \partial_t D_0, \qquad \Gamma = \partial_u D_0
\end{equation} for some potential $D_0$.  

\subsection{Summary of System}

We now summarize the system.  The metric is 
\begin{align}
ds^2_{10} &= \frac{ H^{-1/2}}{\sqrt{y} F_0} \left[ ds^2(AdS_4) +  H e^{2W} \left(dx_1^2 + dx_2^2 \right) + \frac{1}{4} H ds^2_4 \right] \\
ds^2_4 &= g_{ij} di dj + 4 h^{ij} \eta_i \eta_j
\end{align} where $(i,j)$ take value in $(t,u)$.  The metric components are given by
\begin{equation}
g_{ij} = - \partial_i \partial_j D_0, \quad h_{ij} = - \partial_i \partial_j \widetilde{D}_0, \quad h^{ij}h_{jk} = \delta^i_k, \quad \widetilde{D}_0 = D_0 + \frac{4}{3} t (\ln t -1). 
\end{equation} 
We now see that (\eqref{dets}) defines determinants in the internal manifold, $g=\text{det}(g)$ and $h=\text{det}(h)$. When expanded, the metric on the internal manifold $N_4$ is
\begin{equation}
ds^2_4 = -\frac{\det(g)}{\partial_u^2 D_0} dt^2 - \partial_u^2 D_0 \left(du + A_u dt \right)^2 - \frac{4}{\partial_u^2 D_0} \eta_u^2 - \frac{4\partial_u^2 D_0}{\det(h)} \left(\eta_t - A_u \eta_u \right)^2
\end{equation}
where
\begin{equation}
 A_u = \frac{\partial_t \partial_u D_0}{\partial_u^2 D_0}, \qquad \eta_i = d\phi_i + \frac{1}{2} \star_2 d_2 \partial_i D_0.
\end{equation}
We can change to the +/- coordinates used throughout much of the body of this paper via
\begin{equation}
t = a_+ t^+ + a_- t^-, \qquad u = t^+-t^-, \qquad \eta_t = \eta_+ + \eta_-, \qquad \eta_u = a_- \eta_+ - a_+ \eta_-.
\end{equation}
The metric takes the same basic form, shown in equation (\ref{pmmetric}).

For the volume-forms and fluxes, the $(t,u)$ coordinates remain more useful. The volume-form for the internal manifold is
\begin{align}
\Vol(M_6) &=\frac{1}{4\sqrt{3t}} H^{5/2} e^{2W} dx_1\wedge dx_2 \wedge du \wedge \eta_u \wedge dt \wedge \eta_t\\
e^{8A-2\phi} \Vol(M_6) &= \frac{1}{4y F_0^2 } H e^{2W} dx_1\wedge dx_2 \wedge du \wedge \eta_u \wedge dt \wedge \eta_t.
\end{align}
and the four-form flux is given as
\begin{equation}
F_0 F_4 =  \left[J - \frac{1}{2} dt \wedge \left(\eta_t -A_u \eta_u\right) \right] \wedge K - \frac{1}{2y} H J\wedge J,
\end{equation} where 
\begin{align}
J &= e^{2W} dx_1 \wedge dx_2 - \frac{1}{2} \left(du + A_u dt \right) \wedge \eta_u \\
K &= \frac{1}{2y} \left(1- 3t H \right) d \left[ \eta_t - A_u \eta_u\right]-\frac{1}{2} d \left(\frac{3t}{y} H\right) \wedge \left(\eta_t -A_u \eta_u \right).
\end{align}

%%%%%%%%%%%%%%%%%%%%%%%%%%%%%%%%%%%%%%%%%%%%%%%%%%%%%%%%%%%%%%

\section{Kappa-symmetry and BPS Bounds}\label{kappasym}

The requirement for $\kappa$-symmetry for a probe D$p$-brane amounts to the projection \cite{Martucci:2011dn}:
\begin{equation}\label{kappaproj}
\epsilon \equiv \Gamma_{{\rm D}p} \epsilon,
\end{equation}
where $\epsilon$ is defined as a doublet 
\begin{equation}
\epsilon = \begin{pmatrix} \epsilon_1 \\ \epsilon_2 \end{pmatrix}.
\end{equation}
of the supersymmetry parameters $\epsilon_1$, $\epsilon_2$, which are 
Majorana--Weyl spinors of positive and negative chirality respectively.
Furthermore,
\begin{equation}
\Gamma_{{\rm D}p} = 
\begin{pmatrix} 0 & \hat{\Gamma}_{{\rm D}p} \\ \hat{\Gamma}^{-1}_{{\rm D}p} & 0 \end{pmatrix},
\end{equation}
with $\hat{\Gamma}^{-1}_{{\rm D}p} = - (-1)^p \Gamma^0 \hat{\Gamma}^t_{{\rm D}p} \Gamma^0$.
For zero world-volume fluxes, which is the case of interest in this paper,
\begin{align}
\hat{\Gamma}_{{\rm D}p} 
&\equiv \frac{1}{\sqrt{-\det g|_{{\rm D}p}}} \epsilon^{M_1 \dots M_{p}} \Gamma_0 \Gamma_{M_1 \dots M_{p}}|_{{\rm D}p},
\end{align}
where $|_{{\rm D}p}$ denotes the pull-back on the world-volume of the
D$p$-brane. It follows that
\begin{equation}
\hat{\Gamma}^{-1}_{{\rm D}p} = - (-1)^{p(p+1)/2} \hat{\Gamma}_{{\rm D}p}.
\end{equation}

For $\mathcal{N}=2$ supersymmetric AdS$_4 \times M_6$ backgrounds 
the supersymmetry parameters are decomposed as \cite{Passias:2018zlm}
\begin{subequations}
\begin{align}
\epsilon_1 &= \sum_{I = 1}^2 \chi_+^I \otimes \eta_{1+}^I + \sum_{J = 1}^2 \chi_-^J \otimes \eta_{1-}^J,
\\
\epsilon_2 &= \sum_{I = 1}^2 \chi_+^I \otimes \eta_{2-}^I + \sum_{J = 1}^2 \chi_-^J \otimes \eta_{2+}^J,
\end{align}
\end{subequations}
where $\chi^I_\pm$ are AdS$_4$ Killing spinors. A $\pm$ subscript denotes
the chirality of the spinors and the $I,J = 1,2$ indices are R-symmetry
indices. The negative chirality spinors are complex conjugates of the positive chirality ones.
The ten-dimensional gamma matrices are decomposed as 
\begin{equation}
\Gamma_\mu = e^A \gamma_\mu \otimes \mathbb{I} , \qquad
\Gamma_{m + 3} = \gamma_5 \otimes \hat{\gamma}_m , \qquad
\end{equation}
where $\gamma_5 = - i \gamma^0 \gamma^1 \gamma^2 \gamma^3$. 

The spinor bilinears are expressed in terms of the SU(2)-structure data via
the pure spinors $\phi_+^{\pm\pm} \equiv \eta^\pm_{1+}\overline{\eta^\pm_{2+}}$, $\phi_-^{\pm\pm} \equiv \eta^\pm_{1+}\overline{\eta^\pm_{2-}}$,
where $\eta_{i+}^\pm$, $i=1,2$ are defined via
\begin{align}
\eta_{i+}^1 = \frac{1}{\sqrt{2}}(\eta_{i+}^+ + \eta_{i+}^-) \qquad
\eta_{i+}^2 = - \frac{i}{\sqrt{2}}(\eta_{i+}^+ - \eta_{i+}^-).
\end{align}
The pure spinors can be read from equations (2.26) of \cite{Passias:2018zlm},
with the restriction $b=0$, $a=-1$, $z_3=-1$ which applies to the
class of solutions we are studying in this paper.

From the projection \eqref{kappaproj} we can derive the BPS bound 
$\epsilon^\dagger \epsilon \geq \epsilon^\dagger \Gamma_{{\rm D}p} \epsilon$
which can be rewritten as
\begin{align}\label{BPS}
(\epsilon_1^\dagger \epsilon_1 + \epsilon_2^\dagger \epsilon_2) \widetilde{{\rm vol}}_p &\geq U_{(p)},
\end{align}
where
\begin{equation}\label{Up}
\left. U_{(p)} \equiv \left(\epsilon_1^\dagger \Gamma_0 \Gamma_{(p)}\epsilon_2  - (-1)^{p(p+1)/2}  \epsilon_2^\dagger \Gamma_0 \Gamma_{(p)}\epsilon_1 \right)  \right|_{{\rm D}p},
\end{equation}
$\Gamma_{(p)} \equiv \frac{1}{p!} \Gamma_{m_1 \dots m_n} dx^{m_1} \wedge \dots \wedge dx^{m_p}$, and $\widetilde{{\rm vol}}_p$ is the volume form on the spatial part of the world-volume of the D$p$-brane.
The BPS bound is saturated if and only the probe D$p$-brane is supersymmetric.

For the case of a D2-brane that moves along a geodesic in AdS$_4$ and wraps an internal two-cycle we find: 
\begin{equation}
\left. U_{(2)} = 8 e^A {\rm Re}[(\chi_+^1)^\dagger \gamma_0 \chi_-^2]\left(\tfrac{1}{2} dy \wedge (d\psi + \rho)- \frac{e^{-4A}}{F_0^2} J\right) \right|_{{\rm D}2},
\end{equation}
and  
\begin{equation}
(\epsilon_1^\dagger \epsilon_1 + \epsilon_2^\dagger \epsilon_2) \widetilde{{\rm vol}}_2 =
4 e^{A}  \sum_I |\chi_+^I|^2 \widetilde{{\rm vol}}_2.
\end{equation}
We thus conclude that the BPS bound is saturated when
\begin{equation}
\left. \widetilde{{\rm vol}}_2 = \left(\frac{1}{2} dy \wedge (d\psi + \rho)- \frac{e^{-4A}}{F_0^2} J\right)\right|_{{\rm D}2}
\end{equation}
and
\begin{equation}
\sum_I |\chi_+^I|^2 = 2 {\rm Re}[(\chi_+^1)^\dagger \gamma_0 \chi_-^2]
\end{equation}
The latter condition implies that the D2-brane is at the center of AdS$_4$ \cite{Bah:2014dsa}.

%\nocite{*}
\newpage 
\bibliographystyle{utphys}
\bibliography{bibliography5D}

\end{document}